\documentclass[reqno,12pt]{amsart}

\usepackage{amssymb}
\usepackage{fullpage}

\usepackage{amsthm}

\usepackage{amsmath}

\usepackage{amsfonts}

\usepackage{latexsym}

\numberwithin{equation}{section}

{\bf}{\it}

{\bf}{\it}

\newtheorem{theorem}{Theorem}[section]{\bf}{\it}

\newtheorem{lemma}[theorem]{Lemma}{\bf}{\it}

{\bf}{\it}

\newtheorem{proposition}[theorem]{Proposition}{\bf}{\it}

\newtheorem{remark}[theorem]{Remark}{\bf}{\it}

{\bf}{\it}

\def\beq{\begin{equation}}
\def\eeq{\end{equation}}

\def\bal{\begin{align}}
\def\eal{\end{align}}

\newcommand{\ov}{\overline}

\newcommand{\mb}{\mathbb}

\newcommand{\mk}{\mathfrak}

\newcommand{\p}{\partial}

\newcommand{\ha}{\tfrac{1}{2}}

\newcommand{\tb}{\textbf}

\newcommand{\ad}{{\rm ad}}

\def\pa{\parallel}

\def\u{\mathrm{u}}
\def\v{\mathrm{v}}
\def\mm{\mathrm{m}}

\def\hh{\mathrm{h}}
\def\pp{\mathrm{p}}
\def\qq{\mathrm{q}}

\def\ww{\mathrm{w}}
\def\aa{\mathrm{a}}
\def\bb{\mathrm{b}}

\def\tbu{\boldsymbol{\rm u}}
\def\tbw{\boldsymbol{\rm w}}
\def\tbm{\boldsymbol{\rm m}}
\def\tbh{\boldsymbol{\rm h}}

\def\tbs{\boldsymbol{\rm 0}}

\def\tbHH{\boldsymbol{\rm H}}
\def\tbC{\boldsymbol{\rm C}}
\def\ee{\rm e}

\def\<{\langle}
\def\>{\rangle}

\def\e{e}

\def\conx{\omega}

\def\downindex#1{{\mathstrut}_{#1}}

\def\equivH{H^*_\parallel}
\def\ehook#1{\e\downindex{#1}}

\def\Jop{{\mathcal J}}
\def\Hop{{\mathcal H}}

\def\Kop{{\mathcal K}}
\def\Nop{{\mathcal N}}

\def\Ad{{\rm Ad}}

\def\map{\gamma}

\def\inv{{}^{-1}}

\def\Dx{D\downindex{x}}

\def\Dinvx{\Dx^{-1}}

\def\t{{\rm t}}

\def\g{{\rm g}}
\def\A{{\rm A}}
\def\B{{\rm B}}

\def\a{{\rm a}}
\def\b{{\rm b}}
\def\m{\mk{m}}
\def\h{\mk{h}}
\def\Q{\mb{Q}}

\begin{document}

\title{Quaternionic Soliton Equations from Hamiltonian\\ Curve Flows in $\mb{HP}^n$}

\author{
Stephen C. Anco
\lowercase{\scshape{and}}
Esmaeel Asadi\\
\\\lowercase{\scshape{
Department of Mathematics, Brock University,
St. Catharines, ON Canada}}
}

\email{sanco@brocku.ca}
\email{easadi@brocku.ca}

\thanks{S.C.A. is supported by an NSERC research grant. 
The authors thank Takayuki Tsuchida for valuable remarks on parts of this paper.}

\begin{abstract}
A bi-Hamiltonian hierarchy of quaternion soliton equations is derived from
geometric non-stretching flows of curves in the quaternionic projective space
$\mb{HP}^n$.
The derivation adapts the method and results in recent work by one of us
on the Hamiltonian structure of non-stretching curve flows 
in Riemannian symmetric spaces $M=G/H$ by viewing 
$\mb{HP}^n \simeq {\rm U}(n+1,\mb{H})/{\rm U}(1,\mb{H})\times {\rm U}(n,\mb{H})
\simeq {\rm Sp}(n+1)/{\rm Sp}(1)\times {\rm Sp}(n)$
as a symmetric space in terms of compact real symplectic groups
and quaternion unitary groups. 
As main results, scalar-vector (multi-component) versions of
the sine-Gordon (SG) equation 
and the modified Korteweg-de Vries (mKdV) equation
are obtained along with their bi-Hamiltonian integrability structure 
consisting of a shared hierarchy of quaternionic symmetries 
and conservation laws generated by a hereditary recursion operator.
The corresponding geometric curve flows in $\mb{HP}^n$
are shown to be described by a non-stretching wave map 
and a mKdV analog of a non-stretching Schr\"odinger map. 
\end{abstract}
\maketitle

\section{Introduction and Summary}

Certain geometric flows of curves 
in homogeneous plane and space geometries 
are well-known to encode scalar soliton equations 
through the induced evolution of geometrical invariants of the curve 
\cite{Hasimoto,Lamb,GoldsteinPetrich,DoliwaSantini,Pinkall,ChouQu1,ChouQu2,Mari-BeffaSandersWang,Mari-Beffa08,Mari-Beffa09}.
There has been much recent interest in extending such geometrical derivations
to multi-component soliton equations. 
For example, 
the two vector versions of the modified Korteweg-de Vries (mKdV) equation
known from symmetry-integrability classifications \cite{SokolovWolf}
have been derived \cite{AncoSIGMA} from non-stretching curve flows 
in the homogeneous Riemannian geometries 
$M=SO(N+1)/SO(N) \simeq S^N$ and $M=SU(N)/SO(N)$,
generalizing a similar derivation of the scalar mKdV equation
obtained previously \cite{AncoJPHYSA} in the case of the standard 2-sphere geometry,
$S^2 \simeq SO(3)/SO(2) \simeq SU(2)/SO(2)$. 
The same approach has also been used to provide a geometric origin 
for the scalar sine-Gordon (SG) equation 
and its two known vector versions \cite{AncoWolf,AncoSIGMA}.
Related work on integrable curve flows 
in Euclidean and spherical Riemannian geometries, $\mb{R}^N$ and $S^N$,
has appeared in 
\cite{LangerPerline,LangerPerline00,SandersWang,WoQu,Mari-Beffa}.

The derivation of vector mKdV and SG equations in the geometries $M=G/SO(N)$ 
is based on a moving parallel frame formulation 
for arclength-parameterized curves $\gamma$ 
which is closely analogous to the standard parallel framing of curves 
in Euclidean geometry \cite{Bishop,LangerPerline,SandersWang}.
In this generalization \cite{AncoSIGMA}, 
the Euclidean group is replaced by 
the respective isometry groups $G=SO(N+1)$ and $G=SU(N)$ 
whose rotation subgroups $SO(N) \subset G$ act as the gauge group
for the frame bundle of $M=G/SO(N)$ (analogously to that of Euclidean space). 
Given a parallel framing, 
where $x$ is the $G$-invariant arclength along $\gamma$
and $T=\gamma_x$ is the unit tangent vector, 
the Cartan structure equations for 
torsion and curvature of the Riemannian connections $\nabla_x$, $\nabla_t$
on the 2-dimensional surface of any non-stretching curve flow $\gamma(t,x)$
in the manifold $M=G/SO(N)$ 
are seen \cite{AncoSIGMA} to geometrically encode a pair of 
compatible Hamiltonian operators.
This bi-Hamiltonian structure generates a hierarchy of integrable curve flows
in which the frame components of the principal normal vector $N=\nabla_x T$
satisfy vector evolution equations related by a hereditary recursion operator.
Each evolution equation displays invariance under the rotation group
$SO(N-1)$ (in the $SO(N)$ gauge group) that 
acts in the normal space at each point on the curve $\gamma$ 
and preserves the form of the connection matrix of the parallel frame, 
so thus the principal normal components have the meaning of 
geometrical covariants that are determined by $\gamma$
up to $SO(N-1)$ gauge freedom 
(which represents the equivalence group of the framing). 
The lowest-order flow in the respective hierarchies 
in $M=SO(N+1)/SO(N) \simeq S^N$ and $M=SU(N)/SO(N)$
yields two different vector mKdV equations which are $SO(N-1)$-invariant. 
In addition, each hierarchy also contains a flow that yields a
$SO(N-1)$-invariant vector SG equation 
arising from the kernel of the hereditary recursion operator. 
The corresponding geometric curve flows $\gamma(t,x)$ in both hierarchies 
are found to be given by \cite{AncoJPHYSA,AncoSIGMA}
wave maps and mKdV analogs of Schr\"odinger maps on $M=G/SO(N)$. 

Complex versions of these vector SG equations and vector mKdV equations
have been derived \cite{AncoIMA} through applying the same method to 
geometric non-stretching curve flows in the Lie groups 
$G=SO(N+1),SU(N)$ themselves viewed as homogeneous Riemannian geometries 
in the standard manner $G\simeq G\times G/{\rm diag}(G\times G)$. 
This approach has also yielded the two known vector versions of
the nonlinear Schr\"odinger equation (NLS), along with their bi-Hamiltonian
integrability structure. 

A broad generalization of such results has been obtained in recent work 
\cite{AncoJGP} by one of us 
on the Hamiltonian structure of non-stretching flows of curves
in homogeneous Riemannian geometries (i.e. symmetric spaces) $M=G/H$,
including compact semisimple Lie group geometries $M=K$ 
for $G=K\times K$, $H={\rm diag}\ G$.
For all these geometries, 
the results give a general geometric derivation of 
group-invariant (multi-component) SG, mKdV, and NLS equations
along with their bi-Hamiltonian integrability structure consisting of
a shared hierarchy of symmetries and conservation laws 
generated by a group-invariant recursion operator.

In the present paper we adapt this derivation to non-stretching curve flows 
in the quaternion projective geometry $M=\mb{HP}^n$
by means of the symmetric space isomorphisms 
\begin{equation}
\mb{HP}^n \simeq {\rm U}(n+1,\mb{H})/{\rm U}(1,\mb{H})\times {\rm U}(n,\mb{H})
\simeq {\rm Sp}(n+1)/{\rm Sp}(1)\times {\rm Sp}(n) 
\label{HPnsymmstructure}
\end{equation}
given in terms of Hamilton's quaternions $\mb{H}={\rm span}(1,i,j,k)$.
Underlying these isomorphisms is a basic identification \cite{Chevalley}
between the quaternion unitary Lie algebra
$\mk{u}(n,\mb{H})$
and the compact symplectic real Lie algebra $\mk{sp}(n)$
holding for $n\geq 1$. 
Our main result will be to obtain 
bi-Hamiltonian geometric curve flows in $M=\mb{HP}^n$
yielding an integrable multi-component quaternionic version of 
the mKdV equation and the SG equation. 

A summary of relevant properties of 
symplectic groups, quaternion unitary groups, 
and the symmetric space structure of $\mb{HP}^n$ is provided in section~2. 
In section~3, 
starting from the gauge group ${\rm U}(1,\mb{H})\times {\rm U}(n,\mb{H})$ 
of the frame bundle of $\mb{HP}^n$,
we extend the moving parallel frame formulation to curves in $\mb{HP}^n$
and explain its main geometrical properties. 
In particular, 
the components of the principal normal vector along any 
arclength-parameterized curve with this framing 
define geometrical covariants that transform as a quaternion scalar-vector pair
(or a single quaternion scalar in the case $n=1$)
with respect to the equivalence group 
${\rm U}(1,\mb{H})\times {\rm U}(n-1,\mb{H})$ 
(which preserves the form of the parallel connection matrix) of the framing.
In section~4 
we next show how the Cartan structure equations for framed curve flows
encode two compatible quaternion Hamiltonian operators. 
We then use these operators in section~5 
to generate a hierarchy of quaternion Hamiltonian vector fields
and corresponding quaternion scalar-vector evolution equations
with an explicit bi-Hamiltonian structure. 
This yields a quaternion scalar-vector mKdV equation
and a quaternion scalar-vector SG equation, 
both of which are unitarily invariant 
under ${\rm U}(1,\mb{H})\times {\rm U}(n-1,\mb{H})$
and share a hierarchy of quaternionic symmetries and conservation laws 
which are generated by a 
${\rm U}(1,\mb{H})\times {\rm U}(n-1,\mb{H})$-invariant recursion operator.
In section~6
we derive the corresponding geometric SG and mKdV curve flows in $\mb{HP}^n$
and show that these flows respectively are described by 
a non-stretching wave map and a non-stretching mKdV map. 
Finally, we make some concluding remarks in section~7.

\section{Algebraic and Geometric Preliminaries}

Here we collect some useful facts about the relevant algebraic and geometric structure of $\mb{HP}^n$ as a quaternion symmetric space.
 
The complex symplectic group ${\rm Sp}(n,\mb{C})$ is the group of matrices $g$ 
in ${\rm GL}(2n,\mb{C})$ that leaves invariant the exterior form  
$z_1\wedge z_{n+1}+\cdots+z_n\wedge z_{2n}$ in terms of coordinates
$(z_1,\ldots,z_{2n})\in\mb{C}^{2n}$, i.e.
$$ 
g^\t J_ng=J_n,\quad 
J_n=\begin{pmatrix}0&I_n\\-I_n&0\end{pmatrix} .
$$
Recall that the group of matrices $g$ in ${\rm GL}(2n,\mb{C})$ 
that leaves invariant the Hermitian form 
$z_1\overline{z}_1+\cdots+z_{2n}\overline{z}_{2n}$ 
is the complex unitary group ${\rm U}(2n)$, i.e.
$$
g^\t\overline{g}=I_{2n}.
$$
The compact {\it symplectic group}  is defined by ${\rm Sp}(n)={\rm Sp}(n,\mb{C})\cap {\rm U}(2n)$, 
whose elements  have the form
\beq\label{generic.sp}
g=\begin{pmatrix}A&B\\-\overline{B}&\overline{A}\end{pmatrix},\quad A^\t\ov{A}+\ov{B}^\t B=I_n,\quad A^\t\ov{B}-\ov{B}^\t A=0.
\eeq

In the case $n=1$, the compact symplectic group ${\rm Sp}(1)$ has 
a well-known identification with Hamilton's quaternions.
The quaternions are defined as 
an associative non-commutative algebra $\mb{H}$ of dimension $4$ over the real numbers, 
with generators  $i,j,k$ that satisfy the multiplicative relations 
$$ i^2=j^2=k^2=-1,\quad ij=-ji=k,\quad jk=-kj=i,\quad ki=-ik=j. $$
Quaternions $q \in \mb{H}={\rm span}(1,i,j,k)$ have two main properties:
First, 
every $q$ has a natural complex conjugate defined by the conjugation relations
$$ \ov{i}=-i,\quad\ov{j}=-j,\quad\ov{k}=-k,\quad \ov{1}=1, $$
whereby 
$$ \ov{q_1 q_2} = \ov{q}_2 \ov{q}_1 .$$
Second, 
each non-zero $q$ has a multiplicative inverse given by  
$q^{-1}=\bar{q}/|q|^2$, where $|q|=(q\bar{q})^{1/2}$ is the norm of $q$.
Thus, $\mb{H}$ is a normed, hypercomplex, division algebra. 

\begin{proposition}\label{group.prop}
There is a group isomorphism
\beq\label{map.id}
g=\begin{pmatrix}A&B\\-\overline{B}&\overline{A}\end{pmatrix}=
\begin{pmatrix}A_1+A_2i&B_1+B_2i\\-B_1+B_2i&A_1-A_2i\end{pmatrix}\longleftrightarrow  \tilde{g}=A+Bj=A_1+A_2i+B_1j+B_2k
\eeq
under which the compact symplectic group ${\rm Sp}(n)$ is identified with 
the quaternionic unitary group ${\rm U}(n,\mb{H})$ 
given by all matrices $\tilde{g}\in {\rm GL}(n,\mb{H})$ 
satisfying $\tilde{g}^\t\ov{\tilde{g}}=I_{n}$.
In particular, ${\rm Sp}(1)$ is identified with the group formed by unit-norm elements in $\mb{H}$,
\beq\label{g.group.2}
g=\begin{pmatrix}a_1+a_2i&b_1+b_2i\\-b_1+b_2i&a_1-a_2i\end{pmatrix}\longleftrightarrow \tilde{g}=a_1+a_2i+b_1j+b_2k
\eeq
where $a_1^2+a_2^2+b_1^2+b_2^2=1=|\tilde{g}|^2$.
\end{proposition}

The group ${\rm U}(n,\mb{H})\simeq {\rm Sp}(n)$ acts on $\mb{H}^n$ 
by right multiplication, i.e. for $\tb{x}\in\mb{H}^n$ and $\tilde{g}\in{\rm U}(n,\mb{H})$, $\tb{x}\longmapsto \tb{x}\tilde{g}\in\mb{H}^n$. 
There is a natural Hermitian inner product on $\mb{H}^n$ defined by 
$$ <\tb{x},\tb{y}>:=\tb{x}\ov{\tb{y}}^\t $$ 
which is invariant under the group action of ${\rm U}(n,\mb{H})$, i.e.  
$<\tb{x}\tilde{g},\tb{y}\tilde{g}>=<\tb{x},\tb{y}>$ for all $\tilde{g}\in{\rm U}(n,\mb{H})$.
This inner product is the sum of a Euclidean inner product given by
\beq\label{euc.inn}
{\rm Re}<\tb{x},\tb{y}>=\ha<\tb{x},\tb{y}>+\ha<\tb{y},\tb{x}>=\ha(\tb{x}\ov{\tb{y}}^\t+\tb{y}\ov{\tb{x}}^\t)
\eeq
and an imaginary skew-form given by
\beq\label{imag.inn}
{\rm Im}<\tb{x},\tb{y}>=\ha<\tb{x},\tb{y}>-\ha<\tb{y},\tb{x}>=\ha(\tb{x}\ov{\tb{y}}^\t-\tb{y}\ov{\tb{x}}^\t).
\eeq
With respect to the Euclidean inner product, 
there is an orthogonal decomposition of $\mb{H}^n$ into real and imaginary parts
$\mb{H}^n=\mb{R}^n\oplus\Q^n$
where
\beq\label{imaginery.q}
\Q:=\{q\in\mb{H}\,\,|\,\, q+\ov{q}=0\}={\rm span}(i,j,k)
\eeq
will denote the set of imaginary quaternions. 

\begin{proposition}\hfil \newline
1.
The Lie algebra $\mk{g}=\mk{sp}(n)$ of  $G={\rm Sp}(n)$ 
consists of all matrices $\g\in\mk{gl}(2n,\mb{C})$ satisfying
$$ \g J_n+J_n\g^\t=0,\quad \g+\ov{\g}^\t=0. $$
2. 
The group isomorphism ${\rm Sp}(n)\simeq {\rm U}(n,\mb{H})$ induces a Lie algebra isomorphism between the symplectic Lie algebra $\mk{sp}(n)$ and the quaternionic unitary Lie algebra $\mk{u}(n,\mb{H})$ consisting of all matrices $\tilde{\g}\in\mk{gl}(n,\mb{H})$ satisfying $\tilde{\g}+\ov{\tilde{\g}}^\t=0$.
This isomorphism is explicitly given by
\beq\label{map.id.lie}
\g=\begin{pmatrix}\A&\B\\-\overline{\B}&\overline{\A}\end{pmatrix}=
\begin{pmatrix}\A_1+\A_2i&\B_1+\B_2i\\-\B_1+\B_2i&\A_1-\A_2i\end{pmatrix}\longleftrightarrow  \tilde{\g}=\A+\B j=\A_1+\A_2i+\B_1j+\B_2k
\eeq
where $\B=\B^\t,\A+\ov{\A}^\t=0$.
In particular, the Lie algebra  $\mk{sp}(1)$ is identified with  
$\mk{u}(1,\mb{H})=\Q$.\newline
3. 
The Cartan-Killing form on $\mk{g}=\mk{sp}(n)\simeq \mk{u}(n,\mb{H})$ is given by
\beq\label{killing.expl}
\<\g_1,\g_2\>={\rm tr}(\ad(g_1)\ad(g_2))=2(n+1){\rm tr}(\g_1\g_2)=4(n+1){\rm Re}({\rm tr}(\tilde{\g}_1\tilde{\g}_2))=\<\tilde{\g}_1,\tilde{\g}_2\>
\eeq
for $\g_1\leftrightarrow \tilde{\g}_1,\g_2\leftrightarrow \tilde{\g}_2$ 
given by (\ref{map.id.lie}).
In particular, the negative-definite Cartan-Killing norm is explicitly given by
\beq\label{norm.kill}
\<\g,\g\>_{\mk{sp}(n)}=-4(n+1){\rm tr}(\A\ov{\A}^\t+\B\ov{\B}^\t)=-4(n+1){\rm tr}(\A_1\A_1^\t+\A_2\A_2^\t+\B_1\B_1^\t+\B_2\B_2^\t)=\<\tilde{\g},\tilde{\g}\>_{\mk{u}(n,\mb{H})}.
\eeq
4.
For the case $n=1$, where 
$\g=\begin{pmatrix}\a_2i&\b_1+\b_2i\\-\b_1+\b_2i&-\a_2i\end{pmatrix}\in\mk{sp}(1)$ 
is identified with $q=\a_2i+\b_1j+\b_2k\in\mk{u}(1,\mb{H})=\Q$, 
the norm is given by
\beq
-\tfrac{1}{8}\<\g,\g\>_{\mk{sp}(1)}=\a_2^2+\b_1^2+\b_2^2=|q|^2=<q,q>.
\eeq
\end{proposition}

The Lie group ${\rm U}(n+1,\mb{H})$ arises geometrically as the isometry group of the quaternionic projective space $\mb{HP}^n$. 
This space consists of the points on the unit sphere in $\mb{H}^{n+1}$ 
with the identification of  pairs of points $x$ and $x\tilde{g}$ 
for every group element $\tilde{g}\in{\rm U}(1,\mb{H})$. 
Since the symmetry group of the Hermitian inner product on $\mb{H}^{n+1}$ is $G={\rm U}(n+1,\mb{H})$, 
then the action of $G$ modulo the action of ${\rm U}(1,\mb{H})$ represents the non-trivial isometries of $\mb{HP}^n$.
Moreover, 
if we consider the origin in $\mb{HP}^n$ as represented by the point $o=(1,0,\ldots,0)\in\mb{H}^{n+1}$, 
then the isotropy subgroup of ${\rm U}(n+1,\mb{H})$ leaving invariant this point $x=o$ is given by $H={\rm U}(1,\mb{H})\times{\rm U}(n,\mb{H})$. 
Hence $\mb{HP}^n$ is a homogeneous quaternionic manifold
\beq\label{hpn.symmetric}
\mb{HP}^n=\frac{{\rm U}(n+1,\mb{H})}{{\rm U}(1,\mb{H})\times {\rm U}(n,\mb{H})}
\eeq
on which $H^*\simeq \Ad({\rm U}(1,\mb{H})\times{\rm U}(n,\mb{H}))$ acts as the isotropy group  at the origin $x=o$, i.e. 
$H^*$ linearly maps the tangent space $T_o\mb{HP}^n$ into itself.
It is also clear that the group element
\beq\label{S.sigma}
S:=\begin{pmatrix}-1&0\\0&I_n\end{pmatrix}\in{\rm U}(n+1,\mb{H})
\eeq
yields an involutive automorphism of ${\rm U}(n+1,\mb{H})$. 
Consequently, through Proposition \ref{group.prop},  $\mb{HP}^n$ has the structure of a symmetric Riemannian space
\beq\label{hpn.symple}
\mb{HP}^n\simeq \frac{{\rm Sp}(n+1)}{{\rm Sp}(1)\times {\rm Sp}(n)}.
\eeq
The quaternionic structure of this space takes the form of a triple of 
linear maps $I,J,K$ on $T_x\mb{HP}^n$ 
having the properties $IJ=K,JK=I,KI=J,I^2=J^2=K^2=-{\rm id}$, 
as canonically associated with the action of the isotropy subgroup 
$\Ad({\rm Sp}(1))\subset H^*$ on $T_o\mb{HP}^n$.
 
At the Lie algebra level, there is a decomposition of $\mk{g}=\mk{u}(n+1,\mb{H})\simeq \mk{sp}(n+1)$  
as a symmetric Lie algebra 
constructed in terms of the involutive automorphism $\sigma(\g)=S\g S$.
The eigenvalues of $\sigma$ are $1$ and $-1$ with corresponding eigenspaces
\beq\label{rep.eigen.h}
\mk{h}:=\mk{u}(1,\mb{H})\oplus\mk{u}(n,\mb{H})\simeq \mk{sp}(1)\oplus\mk{sp}(n),\quad \sigma(\mk{h})=\mk{h}
\eeq
and
\beq\label{rep.eigen.m}
\mk{m}:=\mk{u}(n+1,\mb{H})/\big(\mk{u}(1,\mb{H})\oplus\mk{u}(n,\mb{H})\big)
\simeq 
\mk{sp}(n+1)/\big(\mk{sp}(1)\oplus\mk{sp}(n)\big),\quad \sigma(\mk{m})=-\mk{m}.
\eeq
Therefore, with respect to the Cartan-Killing form on $\mk{g}$,  
$\sigma$ induces an orthogonal decomposition  
given by the direct sum of vector spaces $\mk{g}=\mk{h}\oplus\mk{m}$,  
with Lie bracket relations
\beq\label{ber.rel}
[\mk{h},\mk{h}]\subset \mk{h},\quad [\mk{h},\mk{m}]
\subset 
\mk{m},\quad [\mk{m},\mk{m}]\subset \mk{h}.
\eeq
Moreover, the Lie subalgebra  $\mk{h}$ is identified with the generators of isometries that leave fixed the origin in $\mb{HP}^n$ 
(i.e. yielding the action of $H^*:T_o\mb{HP}^n\to T_o\mb{HP}^n$), 
while the vector space $\mk{m}$ is identified with the generators of isometries that carry the origin $o$ to any point $x\neq o$ in $\mb{HP}^n$.  

\begin{lemma}\hfil \newline
1.
The matrix representation of the vector space $\mk{m}$ and the Lie subalgebra $\mk{h}$ of $\mk{g}$ in $\mk{gl}(n+1,\mb{H})$ is given by
\begin{align}
(\mm,\tbm) &:=
\begin{pmatrix}0&\mm&\tbm\\-\ov{\mm}&0&\tbs\\-\ov{\tbm}^\t&\tbs&\tbs\end{pmatrix}\in\mk{m},
\label{m.h}\\
(\pp,\qq,\tbh,\tbHH) &:=
\begin{pmatrix}\pp&0&\tbs\\0&\qq&\tbh\\\tbs&-\ov{\tbh}^\t&\tbHH\end{pmatrix}=
\begin{pmatrix}\pp&0&\tbs\\0&0&\tbs\\\tbs&\tbs&\tbs\end{pmatrix}+\begin{pmatrix}0&0&\tbs\\0&\qq&\tbh\\\tbs&-\ov{\tbh}^\t&\tbHH\end{pmatrix}\in\mk{h}=\mk{u}(1,\mb{H})\oplus\mk{u}(n,\mb{H}),
\label{m.h.1}
\end{align}
in which $\tbm,\tbh\in\mathbb{H}^{n-1}$ are quaternionic vectors, 
$\mm\in\mathbb{H}$ is a quaternionic  scalar,  
$\pp,\qq\in\Q$ are imaginary quaternions, 
and $\tbHH\in\mk{u}(n-1,\mb{H})$ is a quaternionic matrix.
The bracket relations (\ref{ber.rel}) take the form:
\begin{subequations}
\begin{align}
[(\mm,\tbm),(\pp,\qq,\tbh,\tbHH)] =&
(\mm\qq-\pp\mm-<\tbm,\tbh>\,,\,\mm\tbh-\pp\tbm+\tbm\tbHH)\in\mk{m}
\label{bracket.m.h}\\
[(\mm_1,\tbm_1),(\mm_2,\tbm_2)] =&
(-\mm_1\ov{\mm}_2+\mm_2\ov{\mm}_1-2{\rm Im}<\tbm_1,\tbm_2>\,,\,
-\ov{\mm}_1\mm_2+\ov{\mm}_2\mm_1\,,\nonumber\\&\quad
\,-\ov{\mm}_1\tbm_2+\ov{\mm}_2\tbm_1\,
,\,-\ov{\tbm}^t_1\tbm_2+\ov{\tbm}^t_2\tbm_1)\in\mk{h},
\label{bracket.m.m}\\
[(\pp_1,\qq_1,\tbh_1,\tbHH_1),(\pp_2,\qq_2,\tbh_2,\tbHH_2)] =&
(\pp_1\pp_2-\pp_2\pp_1\,,\,
\qq_1\qq_{2}-\qq_2\qq_{1}-2{\rm Im}<\tbh_1,\tbh_2>\,,\,\nonumber\\&\quad
\qq_1\tbh_2-\qq_2\tbh_1+\tbh_1\tbHH_2-\tbh_2\tbHH_1\,,\,\nonumber\\&\quad
\ov{\tbh}_1^t\tbh_2-\ov{\tbh}_2^t\tbh_1+\tbHH_1\tbHH_2-\tbHH_2\tbHH_1)\in\mk{h}.\label{bracket.h.h}
\end{align}
\end{subequations}
2.
The restriction of Cartan-Killing form on $\mk{g}$ to $\mk{m}$ yields 
a negative-definite inner product 
\beq\label{kill.on.m}
\<(\mm_1,\tbm_1),(\mm_2,\tbm_2)\>=-\chi {\rm Re}(\mm_1\ov{\mm}_2+<\tbm_1,\tbm_2>)
\eeq
where 
\beq\label{chi}
\chi=8(n+2).
\eeq
3.
The rank of $\mk{m}$ is equal to 1.
\end{lemma}

Cartan subspaces of $\mk{m}$ are defined as a maximal abelian subspace 
$\mk{a}\subseteq\mk{m}$, 
having the property that it is the centralizer of its elements, 
$\mk{a}=\mk{m}\cap \mk{c}(\mk{a})$. 
It can be shown \cite{Helgason} that any two Cartan subspaces are isomorphic to one another under some linear transformation in  ${\rm Ad}(H)$ and that the action of ${\rm Ad}(H)$ on any Cartan subspace $\mk{a}$ generates $\mk{m}$. 
Since $\mk{m}=\mk{u}(n+1,\mb{H})/(\mk{u}(1,\mb{H})\oplus\mk{u}(n,\mb{H}))$ 
has  $\textrm{rank}(\mk{m})=1$, 
its Cartan subspaces $\mk{a}$ are one-dimensional.

A choice of basis element defining $\mk{a}={\rm span}(\ee)$ is given by
\beq
\ee:=(1,\tb{0})=\begin{pmatrix}0&1&\tb{0}\\-1&0&\tb{0}\\\tb{0}&\tb{0}&\tb{0}\end{pmatrix}\in\mk{m},
\eeq
which can be readily verified to satisfy  
\beq\label{generate.m}
\mk{m}\cap \mk{c}(\ee)=span(\ee),\quad \mk{m}={\rm span}({\rm Ad}({\rm U}(1,\mb{H})\times{\rm U}(n,\mb{H}))\ee)
\eeq 
where ${\rm Ad}(h)\ee= {\it h} \ee {\it h} ^{-1}$ for all 
$h=\exp((\pp,\qq,\tbh,\tb{H}))\in{\rm U}(1,\mb{H})\times{\rm U}(n,\mb{H})$ 
with $\pp,\qq\in\Q$, $\tbh\in\mb{H}^{n-1}$, $\tb{H}\in\mk{u}(n-1,\mb{H})$. 
Following the notation used in \cite{AncoJGP}, 
let the centralizer subspaces of $\ee$ in $\mk{m}$ and $\mk{h}$ be denoted 
\beq 
\mk{m}_{\pa}:=\mk{c}(\ee)\cap\mk{m},\quad \mk{h}_{\pa}:=\mk{c}(\ee)\cap\mk{h}.
\eeq
The orthogonal complements (perp spaces) of these subspaces 
$\mk{m}_{\pa}$ and $\mk{h}_{\pa}$ with respect to the Cartan-Killing form 
will be denoted by $\mk{m}_{\perp}$ and $\mk{h}_{\perp}$,  
where 
$$ \mk{m}=\mk{m}_{\pa}\oplus\mk{m}_{\perp},\quad 
\mk{h}=\mk{h}_{\pa}\oplus\mk{h}_{\perp}. $$
Their matrix representation is given by
\beq\label{m.pa.pe}
(\mm_{\pa}):=\begin{pmatrix}0&\mm_{\pa}&\tb{0}\\-\mm_{\pa}&0&\tb{0}\\\tb{0}&\tb{0}&\tb{0}\end{pmatrix}\in\mk{m}_{\pa},\quad
(\mm_{\perp},\tb{m}_{\perp}):=\begin{pmatrix}0&\mm_{\perp}&\tb{m}_{\perp}\\\mm_{\perp}&0&\tb{0}\\-\ov{\tb{m}}_{\perp}^\t&\tb{0}&\tb{0}\end{pmatrix}\in\mk{m}_{\perp},
\eeq
and
\beq\label{h.pa.pe.1}
(\hh_{\pa},\tb{H}_{\pa}):=\begin{pmatrix}\hh_{\pa}&0&\tb{0}\\0&\hh_{\pa}&\tb{0}\\\tb{0}&\tb{0}&\tb{H}_{\pa}\end{pmatrix}\in\mk{h}_{\pa},\quad
(\hh_{\perp},\tb{h}_{\perp}):=\begin{pmatrix}\hh_{\perp}&0&\tb{0}\\0&-\hh_{\perp}&\tb{h}_{\perp}\\\tb{0}&-\ov{\tb{h}}_{\perp}^\t&\tb{0}\end{pmatrix}\in\mk{h}_{\perp},
\eeq
where $\mm_{\pa}\in\mb{R}$ is a real quaternion,  
$\hh_{\pa},\mm_{\perp},\hh_{\perp}\in\Q$ are imaginary quaternions, 
$\tb{m}_{\perp},\tb{h}_{\perp}\in\mathbb{H}^{n-1}$ are quaternionic vectors, 
and $\tb{H}_{\pa}\in\mk{u}(n-1,\mb{H})$ is a quaternionic matrix. 
The corresponding decomposition of $\mk{m}\simeq \mb{H}^n$ as a vector space is given by
\beq\label{m.dec.r.q}
\mk{m}_{\pa}\simeq \mb{R},\quad \mk{m}_{\perp}\simeq \Q\oplus\mb{H}^{n-1}.
\eeq

The linear operator $\ad(\ee)$ maps $\mk{h}_{\perp}$ into $\mk{m}_{\perp}$, 
and vice versa. 
In particular,
\beq\label{ad.e.h.pe}
\ad(\ee)(\hh_{\perp},\tb{h}_{\perp})=(-2\hh_{\perp},\tb{h}_{\perp})\in\mk{m}_{\perp},\quad 
\ad(\ee)(\mm_{\perp},\tb{m}_{\perp})=(2\mm_{\perp},-\tb{m}_{\perp})\in\mk{h}_{\perp}.
\eeq
Consequently, 
$\ad(\ee)^2$ is well-defined as a linear mapping of the subspaces $\mk{h}_{\perp}$ and $\mk{m}_{\perp}$ into themselves.

Let $H^*_{\ee}$ be the  subgroup in the  isotropy group $H^*\simeq {\rm Ad}(H)$ given by
${\rm Ad}(h)\ee=\ee$, $h\in H$, 
generated by the Lie subalgebra $\mk{h}_{\pa}\subset \mk{h}$.
The subgroup $H^*_{\ee}$ can be identified with the group
${\rm U}(n-1,\mb{H})\times{\rm U}(1,\mb{H})\subset {\rm U}(n+1,\mb{H})$ 
whose matrix representation is given by
\beq\label{H.star.e}
\begin{pmatrix}a&0&0\\0&a&0\\0&0&A\end{pmatrix},\quad a\in {\rm U}(1,\mb{H}), \quad A\in{\rm U}(n-1,\mb{H}).
\eeq
Here the subgroup ${\rm U}(n-1,\mb{H})$ acts on $\mk{m}$ by right multiplication, 
\beq\label{act.on.A}
\Ad(A)\mm_{\pa}=\mm_{\pa},\quad \Ad(A)\mm_{\perp}=\mm_{\perp},\quad \Ad(A)\tb{m}_{\perp}=\tb{m}_{\perp}A,
\eeq
while ${\rm U}(1,\mb{H})$ has a non-standard action given by
\beq\label{action.on.m}
{\rm Ad}(a)\mm_{\pa}=\mm_{\pa},\quad {\rm Ad}(a)\mm_{\perp}=a\mm_{\perp}a^{-1},\quad {\rm Ad}(a)\tb{m}_{\perp}=a\tb{m}_{\perp}.
\eeq
The corresponding group action on $\mk{h}$ is given by
\beq\label{action.on.h}
{\rm Ad}(a)\hh_{\pa}=a\hh_{\pa}a^{-1},\quad {\rm Ad}(a)\tb{H}_{\pa}=\tb{H}_{\pa},\quad {\rm Ad}(a)\hh_{\perp}=a\hh_{\perp}a^{-1},\quad {\rm Ad}(a)\tb{h}_{\perp}=a\tb{h}_{\perp}.
\eeq

\begin{proposition}
The vector subspaces $\mk{h}_{\perp}$ and $\mk{m}_{\perp}$ are isomorphic under the linear map $\ad(\ee)$,
and consequently they each decompose into a direct sum of vector subspaces 
given by irreducible representations of the group $H^*_{\ee}$ 
on which the linear map $\ad(\ee)^2$ is a multiple of the identity:
\beq\label{ad.e.2}
\ad(\ee)^2(\mm_{\perp},\tbm_{\perp})=(-4\mm_{\perp},-\tbm_{\perp})\in\mk{m}_{\perp},\quad \ad(\ee)^2(\hh_{\perp},\tbh_{\perp})=(-4\hh_{\perp},-\tbh_{\perp})\in\mk{h}_{\perp}.
\eeq
\end{proposition}

The Lie bracket relations on $\mk{m}=\mk{m}_{\pa}\oplus\mk{m}_{\perp}$ and 
$\mk{h}=\mk{h}_{\pa}\oplus\mk{h}_{\perp}$ coming from the structure of $\mk{g}$ as a symmetric Lie algebra (\ref{ber.rel}) consist of
\begin{align}
[\m_\parallel,\m_\parallel] \subseteq \h_\parallel,\quad
[\m_\parallel,\h_\parallel] \subseteq \m_\parallel,\quad
[\h_\parallel,\h_\parallel] \subseteq \h_\parallel,
\label{inclusion.one}\\
[\m_\parallel,\m_\perp] \subseteq \h_\perp,\quad
[\m_\parallel,\h_\perp] \subseteq \m_\perp,\quad
[\h_\parallel,\m_\perp] \subseteq \m_\perp,\quad
[\h_\parallel,\h_\perp] \subseteq \h_\perp. 
\label{inclusion.two}
\end{align}
The only Lie brackets with nontrivial decompositions are
$[\mk{m}_{\perp},\mk{m}_{\perp}]$, 
$[\mk{h}_{\perp},\mk{h}_{\perp}]$, 
$[\mk{m}_{\perp},\mk{h}_{\perp}]$.
To write out all these brackets explicitly, 
we introduce the following commutator and anti-commutator notations. 
For $\aa,\bb\in\Q$ and $\tb{a},\tb{b}\in\mb{H}^{n-1}$, let
\begin{subequations}
\begin{align}
&C(\aa,\bb):=\aa\bb-\bb\aa\in\Q,\quad A(\aa,\bb):=\aa\bb+\bb\aa\in\mb{R},
\label{not.com.anti} \\&
C(\tb{a},\tb{b}):=<\tb{a},\tb{b}>-<\tb{b},\tb{a}>
=\tb{a}\ov{\tb{b}}^\t-\tb{b}\ov{\tb{a}}^\t\in\Q,
\quad
\tb{C}(\tb{a},\tb{b})=\ov{\tb{a}}^\t\tb{b}-\ov{\tb{b}}^\t\tb{a}\in\mk{u}(n-1,\mb{H}),
\label{not.com.vec} \\&
A(\tb{a},\tb{b}):=<\tb{a},\tb{b}>+<\tb{b},\tb{a}>=\tb{a}\ov{\tb{b}}^\t+\tb{b}\ov{\tb{a}}^\t\in\mb{R}.
\label{not.com.mat}
\end{align}
\end{subequations}

\begin{lemma}\hfil \newline
1.
The Lie brackets (\ref{inclusion.one}),(\ref{inclusion.two}) are given by
\begin{subequations}
\begin{align}
&
[(\mm_{1\pa}),(\mm_{2\pa})]=0\in\mk{h}_{\pa},
\label{inclu.1}\\
&
[(\mm_{\pa}),(\hh_{\pa},\tbHH_{\pa})]=0\in\mk{m}_{\pa},
\label{inclu.2}\\
&
[(\hh_{1\pa},\tbHH_{1\pa}),(\hh_{2\pa},\tbHH_{2\pa})]=
(C(\hh_{1\pa},\hh_{2\pa})\,,\,[\tbHH_{1\pa},\tbHH_{2\pa}])\in\mk{h}_{\pa},
\label{inclu.3}\\
&
[(\mm_{\pa}),(\mm_{\perp},\tbm_{\perp})]=(2\mm_{\pa}\mm_{\perp}\,,\,-\mm_{\pa}\tbm_{\perp})\in\mk{h}_{\perp},
\label{inclu.4}\\
&
[(\mm_{\pa}),(\hh_{\perp},\tbh_{\perp})]=(-2\mm_{\pa}\hh_{\perp}\,,\,\mm_{\pa}\tbh_{\perp})\in\mk{m}_{\perp},
\label{inclu.5}\\
&
[(\hh_{\pa},\tbHH_{\pa}),(\mm_{\perp},\tbm_{\perp})]=
(C(\hh_{\pa},\mm_{\perp})\,,\,\hh_{\pa}\tbm_{\perp}-\tbm_{\perp}\tbHH_{\pa})\in\mk{m}_{\perp},
\label{inclu.6}\\
&
[(\hh_{\pa},\tbHH_{\pa}),(\hh_{\perp},\tbh_{\perp})]=(C(\hh_{\pa},\hh_{\perp})\,,\,\hh_{\pa}\tbh_{\perp}-\tbh_{\perp}\tbHH_{\pa})\in\mk{h}_{\perp}.
\label{inclu.7}
\end{align}
\end{subequations}
2.
The remaining Lie brackets are given by the projections
\begin{subequations}
\begin{align}
&[(\mm_{1\perp},\tbm_{1\perp}),(\mm_{2\perp},\tbm_{2\perp})]_{\mk{h}_{\pa}}=
\big(C(\mm_{1\perp},\mm_{2\perp})+\tfrac{1}{2}C(\tbm_{2\perp},\tbm_{1\perp}),
\tbC(\tbm_{2\perp},\tbm_{1\perp})\big)\in\mk{h}_{\pa},
\label{inclu.8}\\
&[(\mm_{1\perp},\tbm_{1\perp}),(\mm_{2\perp},\tbm_{2\perp})]_{\mk{h}_{\perp}}=
\big(\tfrac{1}{2} C(\tbm_{2\perp},\tbm_{1\perp})\,,\,
\mm_{1\perp}\tbm_{2\perp}-\mm_{2\perp}\tbm_{1\perp}\big)\in\mk{h}_{\perp},
\label{inclu.9}\\
&[(\hh_{1\perp},\tbh_{1\perp}),(\hh_{2\perp},\tbh_{2\perp})]_{\mk{h}_{\pa}}=
\big(C(\hh_{1\perp},\hh_{2\perp})
-\tfrac{1}{2}C(\tbh_{2\perp},\tbh_{1\perp})\,,\,
\tbC(\tbh_{2\perp},\tbh_{1\perp})\big)\in\mk{h}_{\pa},
\label{inclu.10}\\
&[(\hh_{1\perp},\tbh_{1\perp}),(\hh_{2\perp},\tbh_{2\perp})]_{\mk{h}_{\perp}}=
\big(\tfrac{1}{2}C(\tbh_{1\perp},\tbh_{2\perp})\,,\,
\hh_{2\perp}\tbh_{1\perp}-\hh_{1\perp}\tbh_{2\perp}\big)\in\mk{h}_{\perp},
\label{inclu.11}\\
&[(\mm_{\perp},\tbm_{\perp}),(\hh_{\perp},\tbh_{\perp})]_{\mk{m}_{\pa}}=
\big(-A(\mm_{\perp},\hh_{\perp})-\tfrac{1}{2}A(\tbm_{\perp},\tbh_{\perp})\big)\in\mk{m}_{\pa},
\label{inclu.12}\\
&[(\mm_{\perp},\tbm_{\perp}),(\hh_{\perp},\tbh_{\perp})]_{\mk{m}_{\perp}}=
\big(\tfrac{1}{2}C(\tbh_{\perp},\tbm_{\perp})\,,\,\mm_{\perp}\tbh_{\perp}-\hh_{\perp}\tbm_{\perp}\big)\in\mk{m}_{\perp}.
\label{inclu.13}
\end{align}
\end{subequations}
3. 
The Cartan-Killing form on $\mk{h}_{\perp}\simeq \mk{m}_{\perp}$ is given by
\beq\label{killing.h}
\<(\hh_{1\perp},\tbh_{1\perp}),(\hh_{2\perp},\tbh_{2\perp})\>=
\chi{\rm Re}(\hh_{1\perp}\hh_{2\perp}-\tbh_{1\perp}\ov{\tbh}^\t_{2\perp})=
\frac{\chi}{2}(A(\hh_{1\perp},\hh_{2\perp})-A(\tbh_{1\perp},\tbh_{2\perp})).
\eeq
\end{lemma}

An explicit basis for $\mk{m}$ can be built from the basis $\{1,i,j,k\}$ for $\mb{R}\oplus\Q$ together with the standard orthonormal basis $\{\tb{e}_l\}_{l=1,\ldots,n-1}$ of $\mb{R}^{n-1}$  as follows:
\beq\label{basis.m}
\begin{aligned}
& \mk{m}_{\pa}={\rm span}(\ee)\simeq \mb{R}, \\
& \mk{m}_{\perp}={\rm span}({\tbm}_{\perp}^{\it i},{\tbm}_{\perp}^{\it j},{\tbm}_{\perp}^{\it k})\oplus{\rm span}({\tbm}_{\perp\it l},{\tbm}_{\perp\it l}^{\it i},{\tbm}_{\perp\it l}^{\it j},{\tbm}_{\perp\it l}^{\it k})_{{\it l}=1,\ldots,n-1}\simeq \Q\oplus\mb{H}^{n-1}
\end{aligned}
\eeq
in which
\beq\label{tbm.l}
\begin{gathered}
\tb{m}_{\perp l}=(0,\tb{e}_l)=\begin{pmatrix}0&0&\tb{e}_l\\0&0&\tb{0}\\-\tb{e}_{l}^\t&\tb{0}&\tb{0}\end{pmatrix},\quad
\tb{m}_{\perp l}^q=(0,q\tb{e}_l)=\begin{pmatrix}0&0&q\tb{e}_l\\0&0&\tb{0}\\\tb{e}_{l}^\t q&\tb{0}&\tb{0}\end{pmatrix},
\\
\tb{m}_{\perp}^q=(q,\tb{0})=\begin{pmatrix}0&q&\tb{0}\\q&0&\tb{0}\\\tb{0}&\tb{0}&\tb{0}\end{pmatrix},
\end{gathered}
\eeq
where $q=i,j,k$, and $l=1,\ldots,n-1$. 
Note ${\tbm}_{\perp\it l}^{\it q}$ and ${\tbm}_{\perp}^{\it q}$ are, respectively, 
related to ${\tbm}_{\perp\it l}$ and $\ee$ by the action of 
${\rm U}(1,\mb{H})\subset {\rm U}(n+1,\mb{H})$ 
defined by ${\rm Ad}(a)\ee={\tbm}_{\perp}^{\it q}$, 
${\rm Ad}(a){\tbm}_{\perp\it l}={\tbm}_{\perp\it l}^{\it q}$
where
\beq\label{a.in.h}
a=\begin{pmatrix}q&0&0\\0&1&0\\0&0&I_{n-1}\end{pmatrix},\quad q=i,j,k\in\Q.
\eeq
In addition, 
$\tb{m}_{\perp l}$ is itself related to $\ee$ 
by the action of the rotation subgroup 
${\rm SO}(n,\mb{R})\subset {\rm U}(n,\mb{H})\subset {\rm U}(n+1,\mb{H})$. 

Finally, we record several useful quaternionic identities connected with properties of the Cartan-Killing form (\ref{killing.h}).
\begin{enumerate}
\item For $a\in\Q$, $b,c\in\mb{H}$ and $\tb{b},\tb{c}\in\mb{H}^{n-1}:$
\begin{align}
&\ov{<b,c>}=<c,b>,\quad 
{\rm Re}(a<b,c>)={\rm Re}(<b,c>a)=-{\rm Re}(a<c,b>),
\label{iden.a}\\
&\ov{<\tb{b},\tb{c}>}=<\tb{c},\tb{b}>,\quad
{\rm Re}(a<\tb{b},\tb{c}>)={\rm Re}(<\tb{b},\tb{c}>a)=-{\rm Re}(a<\tb{c},\tb{b}>).
\label{iden.d}
\end{align}
\item For $a,b,c\in\Q:$
\beq
{\rm Re}(ab)={\rm Re}(ba),\quad {\rm Re}(abc)={\rm Re}(bca)={\rm Re}(cab)=-{\rm Re}(bac)=-{\rm Re}(cba)=-{\rm Re}(acb).
\label{iden.f}
\eeq
\end{enumerate}

\section{Frame Formulation of Non-Stretching Curve Flows in $\mb{HP}^n$}

We consider non-stretching flows of framed curves in the quaternionic projective space $M=\mb{HP}^n$ viewed as a symmetric Riemannian geometry (\ref{hpn.symmetric}). 
Its Riemannian structure will be described in terms of \cite{KobayashiNomizu} a $\mk{m}$-valued linear coframe $e$ and a $\mk{h}$-valued linear connection $\conx$ whose torsion and curvature
\beq\label{tor.curv}
\mk{T}:=de+\tb{[}\conx,e\tb{]},\quad 
\mk{R}:=d\conx+\ha\tb{[}\conx,\conx\tb{]}
\eeq
are $2$-forms with respective values in $\mk{m}$ and $\mk{h}$, 
given by the following Cartan structure equations:
\beq\label{cart.stru}
\mk{T}=0,\quad \mk{R}=-\ha\tb{[}e,e\tb{]}.
\eeq
Here $\tb{[}\cdot,\cdot\tb{]}$ denotes the Lie bracket on $\mk{g}=\mk{u}(n+1,\mb{H})$ composed with the wedge product on $T_x^*M\simeq \mb{H}^n$. 
This structure together with the (negative-definite) Cartan-Killing form 
determines a Riemannian metric and Riemannian connection 
(i.e. covariant derivative) on the manifold $M$ as follows. 
For all $X,Y$ in $T_x M$,
\beq\label{metr.conn}
g(X,Y):=-\<e_X,e_Y\>,\quad e\rfloor\nabla_X Y:=d_Xe_Y+[\conx_X,e_Y],
\eeq
where the coframe provides an identification between the tangent space $T_x M$ and the vector space $\mk{m}=\mk{g}/\mk{h}$ as given by
$e\rfloor X:=e_X,e\rfloor Y:=e_Y\in\mk{m}$. 
The connection is metric compatible, $\nabla g=0$, and torsion-free, $T=0$, while its curvature is covariantly constant, $\nabla  R=0$. 
In particular,
\beq\label{tor.curv.tensor}
e\rfloor R(X,Y)Z=[\mk{R}\rfloor(X\wedge Y),e_Z]=-[[e_X,e_Y],e_Z],\quad e\rfloor T(X,Y)=\mk{T}(X\wedge Y)=0,
\eeq
where $T(X,Y):=\nabla_XY-\nabla_YX-[X,Y]$ is the torsion tensor and $R(X,Y):=[\nabla_X,\nabla_Y]-\nabla_{[X,Y]}$ is the curvature tensor.  
Note the linear coframe and linear connection have gauge freedom 
given by the following transformations
\beq\label{gauge.tr}
e\longrightarrow \Ad(h^{-1})e,\quad\conx\longrightarrow\Ad(h^{-1})\conx+h^{-1}dh
\eeq
for an arbitrary function
$h:M\to H={\rm U}(n,\mb{H})\times {\rm U}(1,\mb{H})\subset {\rm U}(n+1,\mb{H})$. 
These gauge transformations comprise a local ($x$-dependent) representation of the linear isotropy group $H^*=\Ad(H)$ 
which defines the gauge group \cite{Sharpe} of the frame bundle of $M=\mb{HP}^n$. 
Both the metric tensor $g$ and curvature tensor $R$ on $M=\mb{HP}^n$ are gauge invariant.

Let $\gamma(t,x)$ be a flow of any smooth curve in $M=\mb{HP}^n$. 
We write $X=\gamma_x$ for the tangent vector and $Y=\gamma_t$ for the evolution vector at each point $x$ along the curve.  
Note the flow is {\it non-stretching} provided that 
it preserves the ${\rm U}(n+1,\mb{H})$-invariant arclength $ds=|\gamma_x|dx$, 
or equivalently $\nabla_t|\gamma_x|=0$, 
in which case we can put 
\beq\label{non.stret}
|\gamma_x|=1
\eeq
without loss of generality.
For smooth flows that are transverse to the curve (such that $X$ and $Y$ are linearly independent), $\gamma(t,x)$ will describe a smooth two-dimensional surface in $M$. 
The pullback of the Cartan structure equations (\ref{cart.stru}) to this surface yields a framing of the curve evolving under the flow:
\begin{align}
&
D_xe_t-D_te_x+[\conx_x,e_t]-[\conx_t,e_x]=0,
\label{pull.1}\\
&
D_x\conx_t-D_t\conx_x+[\conx_x,\conx_t]=-[e_x,e_t],
\label{pull.2}
\end{align}
with
\begin{align}
&
e_x:=e\rfloor X=e\rfloor \gamma_x,\quad e_t:=e\rfloor Y=e\rfloor \gamma_t,
\label{pull.not.1}\\
&
\conx_x:=\conx\rfloor X=\conx\rfloor \gamma_x,\quad \conx_t:=\conx\rfloor Y=\conx\rfloor \gamma_t,
\label{pull.not.2}
\end{align}
where $D_x,D_t$ denote total derivatives with respect to $x,t$.

Remarkably, 
for any non-stretching curve flow, 
these torsion and curvature equations (\ref{pull.1})--(\ref{pull.not.2}) 
encode an explicit bi-Hamiltonian structure once a specific choice of frame along $\gamma(t,x)$ is made \cite{AncoJGP}. 
A frame consists of a set of orthonormal vectors that span the 
tangent space of $M$ at each point $x$ on the curve. 
Associated to a frame is the connection matrix consisting of the set of frame components of the covariant $x$-derivative of each frame vector along the curve \cite{Guggenheimer}. 
Such a framing is obtained from the Lie-algebra components of  $e^*$ and $\conx_x$ when an orthonormal basis is introduced for $\mk{m}$ and $\mk{h}$ with respect to the Cartan-Killing form, 
where $e^*$ is a $\mk{m}$-valued linear frame 
defined to be dual to the coframe $e$ by the condition that 
$-\<e^*,e\>=\textrm{id}$ 
is the identity map on each tangent space $T_{\gamma}M$ 
(cf \cite{AncoJGP,KobayashiNomizu}).
In particular, $\conx_x$ determines $e$ along the curve via the transport equation
\beq\label{frenet.eq}
\nabla_xe=-\ad(\conx_x)e.
\eeq
Then if $\{\tb{m}_l\}_{l=1,\ldots,4n-1}$ is any fixed orthonormal basis 
for the perp space of ${\rm span}(e_x)$ in $\mk{m}$, 
the set of vectors given by $X_{\perp l}:=-\<e^*,\tb{m}_l\>$ 
together with $X=-\<e^*,e_x\>$ 
defines a frame at each point $x$ along the curve, i.e. 
${\rm span}(X_{\perp 1},\ldots,X_{\perp 4n-1},X)=T_{\gamma}M$ 
with the subset $\{X_{\perp l}\}_{l=1,\ldots,4n-1}$ 
being an orthonormal basis for the normal space of the curve. 
It turns out that the bi-Hamiltonian structure encoded in the resulting frame structure equations for $\gamma(t,x)$ 
is independent of the choice of a basis, 
and accordingly the simplest algebraic formulation of this encoding arises
in terms of the $\mk{m}$-valued coframe and $\mk{h}$-valued connection variables (\ref{pull.not.1}) and (\ref{pull.not.2}).

We utilize a natural choice of a moving frame defined by the following two properties which are a direct algebraic generalization of a parallel moving frame in Euclidean geometry.
\begin{enumerate}
\item[(1)] 
$e_x$ is a constant unit-norm element lying in a fixed Cartan subspace $\mk{a}\subset \mk{m}$ 
(with ${\rm dim}(\mk{a})={\rm rank}(\mk{m})=1$), i.e. 
$D_xe_x=D_te_x=0, \<e_x,e_x\>=-1$, and $\mk{a}={\rm span}(e_x)=\mk{m}_{\pa}$ 
where $\mk{m}_{\pa}\oplus\mk{m}_{\perp}=\mk{m}$ and $\<\mk{m}_{\pa},\mk{m}_{\perp}\>=0$.
\item[(2)] 
$\conx_x$ lies in the perp space $\mk{h}_{\perp}$ of the Lie subalgebra 
$\mk{h}_{\pa}\subset \mk{h}$ of the linear isotropy group 
$H^*_{\pa}\subset H^*\simeq \Ad ({\rm U}(1,\mb{H})\times {\rm U}(n,\mb{H}))$ 
that preserves $e_x$, i.e. 
$\ad(\mk{h}_{\pa})e_x=0$ and $\<\conx_x,\mk{h}_{\pa}\>=0$ 
where
$\mk{h}_{\pa}\oplus\mk{h}_{\perp}=\mk{h}$ and $\<\mk{h}_{\pa},\mk{h}_{\perp}\>=0$.
\end{enumerate}

Such a moving frame for $\gamma(t,x)$ will be called ${\rm U}(1,\mb{H})\times {\rm U}(n,\mb{H})$-{\it parallel} 
and its existence can be established by constructing a suitable gauge transformation (\ref{gauge.tr}) on an arbitrary moving frame at each point $x$ along the curve \cite{AncoJGP}. 
Since every Cartan subspace $\mk{a}\subset\mk{m}$ is one-dimensional, 
a ${\rm U}(1,\mb{H})\times {\rm U}(n,\mb{H})$-parallel moving frame is unique 
up to the rigid ($x$-independent) action of $H^*_{\pa}$ on $e$ and $\conx_x$. 
Specifically, 
given any $\mk{m}$-valued linear coframe $\tilde{e}$ and $\mk{h}$-valued linear connection matrix $\tilde{\conx}_x$ along $\gamma$, 
we can first find a gauge transformation such that 
$h^{-1}\tilde{e}_xh=e_x$ is a constant element in any Cartan subspace 
$\mk{a}\subset\mk{m}$, 
as a consequence of the fact $\mk{m}={\rm Ad}(H)\mk{a}$. 
The norm of $e_x$ will satisfy 
$-\<e_x,e_x\>=g(\gamma_x,\gamma_x)=|\gamma_x|^2=1$ 
because we have chosen an arclength parametrization of the curve. 
We can then find a gauge transformation belonging to the subgroup 
$H^*_{\pa}$ which preserves $e_x$, 
so that $h^{-1}D_xh+h^{-1}\tilde{\conx}_xh=\conx_x$ 
where $h(x)\in H^*_{\pa}$ is given by solving the linear matrix ODE 
$D_xh+\tilde{\varpi}^{\pa}h=0$ 
in terms of the decomposition of $\tilde{\conx}_x=\tilde{\varpi}^{\pa}+\tilde{\varpi}^{\perp}$ relative to $e_x$.
Note the solution will depend on an arbitrary initial condition
$h(x_0)\in H^*_{\pa}$, specified at some point $x=x_0$ along the curve, 
which represents a rigid gauge freedom (i.e. the equivalence group)
in the construction of the moving frame.
 
Employing the quaternionic matrix notation and algebraic preliminaries from Section~2, we choose
\beq\label{mat.con}
\begin{aligned}
&
e_x=\frac{1}{\sqrt{\chi}}\begin{pmatrix}0&1&\tb{0}\\-1&0&\tb{0}\\\tb{0}&\tb{0}&\tb{0}\end{pmatrix}=:
\frac{1}{\sqrt{\chi}}(1,\tb{0})\in\mk{a}\simeq \mb{R},
\\
&
\conx_x=\begin{pmatrix}\u&0&\tb{0}\\0&-\u&\tb{u}\\\tb{0}&-\ov{\tb{u}}^\t&\tb{0}\end{pmatrix}=:
(\u,\tb{u})\in\mk{h}_{\perp}\simeq \Q\oplus\mb{H}^{n-1},
\end{aligned}
\eeq
where  ${\rm u}\in\Q$ is an imaginary quaternion variable, and $\tb{u}\in\mb{H}^{n-1}$ is a quaternionic vector variable. 
(Note the factor $1/\sqrt{\chi}$ in $e_x$ comes from the normalization factor (\ref{chi}) of the Cartan-Killing inner product on $\mk{m}$.) 
The equivalence group $H^*_{\pa}$ of the ${\rm U}(1,\mb{H})\times {\rm U}(n,\mb{H})$-parallel frame obtained from this choice of $e_x$ and $\conx_x$ 
consists of rigid ($x$-independent) gauge transformations (\ref{gauge.tr}) 
that preserve the form of the matrices (\ref{mat.con}). 
Specifically, we have from (\ref{action.on.m}) and (\ref{action.on.h}),  
$\Ad(h^{-1})(1,\tb{0})=(1,\tb{0})$ and $\Ad(h^{-1})(\u,\tb{u})=(\tilde{\u},\tilde{\tb{u}})$
where
\beq\label{gauge.tr.1}
\tilde{\u}=a\u a^{-1}\in \Q\quad\textrm{and}\quad 
\tilde{\tb{u}}=a\tb{u}A\in\mb{H}^{n-1}
\eeq
for all constant functions (on $M$) 
$h=(a,A)\in H_{\pa}={\rm U}(1,\mb{H})\times {\rm U}(n-1,\mb{H})$
with the explicit matrix form (\ref{H.star.e}).

Hereafter $e$ will denote the $\mk{m}$-valued linear coframe along $\gamma(t,x)$ as determined (up to equivalence) by (\ref{mat.con}) through the transport condition  (\ref{frenet.eq}). 
In terms of the dual frame $e^*$, 
the associated framing for $T_{\gamma}M$ will look like
\beq\label{basis.1}
\begin{aligned}
& X:=-\<e^*,e_x\>=\gamma_x,\quad 
X^{}_{\perp l}:=-\<e^*,\tb{m}_{\perp l}\>\perp\gamma_x , \\
& X^q_{\perp}:=-\<e^*,\tb{m}_{\perp}^q\>\perp \gamma_x,\quad 
X^q_{\perp l}:=-\<e^*,\tb{m}_{\perp l}^q\>\perp\gamma_x,\quad q=i,j,k\in\Q
\end{aligned}
\eeq
where 
$\{\tb{m}_{\perp}^q,\tb{m}_{\perp l},\tb{m}_{\perp l}^q\}_{l=1,\ldots,n-1}$ 
give an orthonormal basis for $\mk{m}_{\perp}$ 
while $\{e_x\}$ is a unit basis for $\mk{m}_{\pa}$ 
whence $\{X,X_{\perp}^q,X^{}_{\perp l},X_{\perp l}^q\}_{l=1,\ldots,n-1}$ is an orthonormal frame along $\gamma(t,x)$ such that $dx=ds$ 
is the ${\rm U}(n+1,\mb{H})$-invariant arclength.
An explicit choice of such a basis for $\mk{m}$ is shown in (\ref{tbm.l}). 
The action of $\Ad(H_{\pa})\simeq H^*_{\pa}$ on this basis 
will yield all other choices of an orthonormal basis for $\mk{m}$ 
as given by (\ref{act.on.A})--(\ref{action.on.m}).

The tangent space of $M$ along  $\gamma(t,x)$ thereby has 
an orthogonal decomposition into real and imaginary subspaces  
$T_{\gamma}M\simeq \mb{R}^n\oplus\Q^n$ defined by
\[
\mb{R}^n={\rm span}(X,X^{}_{\perp l})_{l=1,\ldots,n-1}=:{\rm Re} T_{\gamma}M,\quad
\Q^n={\rm span}(X^q_\perp,X^q_{\perp l})_{l=1,\ldots,n-1}^{q=i,j,k}=:{\rm Im} T_{\gamma}M.
\]
In terms of this structure, 
the ${\rm U}(1,\mb{H})\times {\rm U}(n,\mb{H})$-parallel frame (\ref{basis.1}) has the geometrical property
\beq
{\rm Re}\nabla_xX\in{\rm span}(X^{}_{\perp 1},\ldots,X^{}_{\perp n-1})\quad\textrm{and}\quad
{\rm Re}\nabla_xX^{}_{\perp l}\in{\rm span}(X),\quad l=1,\ldots,n-1,
\eeq
closely resembling a Euclidean parallel frame \cite{Bishop}.

The geometrical meaning of the ${\rm U}(1,\mb{H})\times{\rm U}(n,\mb{H})$-parallel connection determined by (\ref{basis.1}) 
is seen from looking at the frame components of the principal normal vector 
\beq\label{prin.no.vec}
N:=\nabla_xX=\<e^*,\ad(e_x)\conx_x\>
\eeq
given by 
\beq\label{ad.om}
e\rfloor N=-\ad(e_x)\conx_x=\frac{1}{\sqrt{\chi}}\begin{pmatrix}0&2\u&-\tb{u}\\
2\u&0&\tb{0}\\
\ov{\tb{u}}^t&\tb{0}&\tb{0}\end{pmatrix}\in\mk{m}_{\perp}.
\eeq
These components $\u$ and $\tb{u}$ are invariantly defined by $\gamma(t,x)$ up to the rigid  ($x$-independent) action of the equivalence group $H^*_{\pa}\simeq \Ad({\rm U}(1,\mb{H})\times{\rm U}(n-1,\mb{H}))$ that preserves the framing at each point $x$. 
Hence, in geometrical terms, 
the scalar-vector pair $(\u,\tb{u})$ describes {\it covariants} of the curve $\gamma$ relative to the group $H^*_{\pa}$, 
such that each component $\u\in\Q$, $\tb{u}\in\mb{H}^{n-1}$ 
belongs to an irreducible representation of this group  
corresponding to the orthogonal decomposition of the vector space 
$\mk{m}_{\perp}\simeq \Q\oplus\mb{H}^{n-1}$.  
Moreover, $x$-derivatives of $(\u,\tb{u})$ describe 
differential covariants of $\gamma$ relative to $H^*_{\pa}$, 
which arise geometrically from the frame components of $\nabla_x$-derivatives of the principal normal vector $N$. 
For example, $(\u_x,\tb{u}_x)$ corresponds to 
\beq\label{ad.om.1}
\begin{aligned}
e\rfloor \nabla_xN &= -\ad(e_x)D_x\conx_x+\ad(\conx_x)^2e_x \\
&= \frac{1}{\sqrt{\chi}}\begin{pmatrix}0&2\u_x+4\u^2-|\tbu|^2&-\tb{u}_x-3\u\tbu\\
2\u_x-4\u^2+|\tbu|^2&0&\tb{0}\\
\ov{\tb{u}}^t_x-3\ov{\tbu}^\t\u&\tb{0}&\tb{0}\end{pmatrix}\in\mk{m}_{\perp}.
\end{aligned}
\eeq

We thus note that the geometric invariants of $\gamma$ 
as defined by Riemannian inner products of the tangent vector $X=\gamma_x$ 
and its derivatives $N=\nabla_x\gamma_x,\nabla_xN=\nabla_x^2\gamma_x$, etc.\ 
along the curve $\gamma$ 
can be expressed as scalars formed from Cartan-Killing inner products of 
the covariants $(\u,\tb{u})$ and differential covariants $(\u_x,\tb{u}_x),(\u_{2x},\tb{u}_{2x})$, etc.; 
in particular,
\begin{gather}
g(N,N)=-g(X,\nabla_x^2X)
=4|\u|^2+|\tb{u}|^2,
\label{nn.g}\\
g(N,\nabla_xN)=-g(X,\nabla_x^3X)
=4{\rm Re}\<\u,\u_x\>+{\rm Re}\<\tb{u},\tb{u}_x\>,
\label{nxn.g}\\
g(\nabla_xN,\nabla_xN)=g(X,\nabla_x^4X)
=4|\u_x|^2+|\tb{u}_x|^2+(4|\u|^2+|\tbu|^2)^2+9|\u|^2|\tbu|^2+6{\rm Re}\<\u\tbu,\tbu_x\>
\label{nxnx.g}
\end{gather}
comprise all ${\rm U}(n+1,\mb{H})$-invariants depending on at most 
$\u,\tb{u},\u_x,\tb{u}_x$. 

\begin{remark}
The set of $4n-1$ invariants given by $\{g(X,\nabla_x^{l}X)\}_{l=2,\ldots,4n}$ 
(and their $x$-derivatives up to differential order $4n-l$)
generate the components of the connection matrix of  a classical Fren\'et frame \cite{Guggenheimer} determined by $\gamma_x$. 
\end{remark}

\section{Bi-Hamiltonian Structure}

Let $\gamma(t,x)$ be any non-stretching curve flow in $M=\mb{HP}^n$ 
and choose a ${\rm U}(1,\mb{H})\times{\rm U}(n,\mb{H})$-parallel framing given by (\ref{mat.con}).  
In the flow direction $Y=\gamma_t$, 
we decompose $e_t=h_{\pa}+h_{\perp}$ relative to $e_x$, with 
\begin{align}
& h_{\pa}=\frac{1}{\sqrt{\chi}}\begin{pmatrix}0&\hh_{\pa}&\tb{0}\\-\hh_{\pa}&0&\tb{0}\\\tb{0}&\tb{0}&\tb{0}\end{pmatrix}=:
\frac{1}{\sqrt{\chi}}(\hh_{\pa})\in\mk{m}_{\pa}\simeq \mb{R},
\label{h.pa.pe}\\
& h_{\perp}=\begin{pmatrix}0&\hh_{\perp}&\tb{h}_{\perp}\\\hh_{\perp}&0&\tb{0}\\-\ov{\tb{h}}_{\perp}^\t&\tb{0}&\tb{0}\end{pmatrix}=:(\hh_{\perp},\tb{h}_{\perp})\in\mk{m}_{\perp}\simeq \Q\oplus\mb{H}^{n-1}, 
\label{hh.pa.pe}
\end{align}
and likewise for $\conx_t=\varpi^{\pa}+\varpi^{\perp}$ with 
\begin{align}
&\varpi^{\pa}=\begin{pmatrix}\ww^{\pa}&0&\tb{0}\\0&\ww^{\pa}&\tb{0}\\\tb{0}&\tb{0}&\tb{W}^{\pa}\end{pmatrix}=:
(\ww^{\pa},\tb{W}^{\pa})\in\mk{h}_{\pa}\simeq \Q\oplus\mk{u}(n-1,\mb{H}),
\label{var.pa.f}\\
& \varpi^{\perp}=
\begin{pmatrix}\ww^{\perp}&0&\tb{0}\\0&-\ww^{\perp}&\tb{w}^{\perp}\\\tb{0}&-\ov{\tb{w}}^{\perp \t}&\tb{0}\end{pmatrix}=:(\ww^{\perp},\tb{w}^{\perp})\in\mk{h}_{\perp}\simeq \Q\oplus\mb{H}^{n-1},
\label{var.pe.f}
\end{align}
where $\hh_{\pa}\in\mb{R}$ is a real scalar variable, 
$\hh_{\perp},\ww^{\pa},\ww^{\perp}\in\Q$ are imaginary quaternion variables, 
$\tb{h}_{\perp},\tb{w}^{\perp}\in\mb{H}^{n-1}$ are quaternion vector variables,
and $\tb{W}^{\pa}\in\mk{u}(n-1,\mb{H})$ is a quaternion unitary matrix. 
Here we have inserted a factor $1/\sqrt{\chi}$ in $h_{\pa}$, which corresponds to the normalization factor in $e_x$, in order to simplify later expressions. 

The frame formulation of the flow $\gamma(t,x)$ is then given by the general results in \cite{AncoJGP} as follows.

\begin{lemma}\label{lemma.cartan.eq}
The Cartan structure equations (\ref{pull.1}),(\ref{pull.2}) for any ${\rm U}(1,\mb{H})\times{\rm U}(n,\mb{H})$-parallel
linear coframe $\e$ and linear connection $\conx$
pulled back to the two-dimensional surface $\map(t,x)$
take the form of  a flow on $\conx_x=:u(t,x)$ given by 
\beq
u_t
= \Dx \varpi^\perp +[u,\varpi^\parallel] + [u,\varpi^\perp]_\perp
+\ad(\e_x) h_\perp ,
\label{uteq}
\eeq
where
\begin{align}
&\varpi^\perp
= -\ad(\e_x)\inv( \Dx h_\perp +[u,h_\parallel] +[u,h_\perp]_\perp ) ,
\label{wperpeq}\\
&h_\parallel = -\Dinvx [u,h_\perp]_\parallel ,\quad
\varpi^\parallel = -\Dinvx [u,\varpi^\perp]_\parallel,
\label{hw.pa.eq}
\end{align}
are given in terms of $h_{\perp}$, 
with $\Dinvx$ denoting the formal inverse of the total $x$-derivative operator $\Dx$.
\end{lemma}

Note that $h_{\perp}$ 
geometrically corresponds to the normal part of the flow vector $Y$, 
$$h_{\perp}=e\rfloor Y_{\perp}$$
where $Y_{\perp}$ is the orthogonal projection of $Y$ relative to the tangent vector $X$ along the curve. 
Similarly, 
the tangential part of the flow vector $Y$ corresponds to $h_{\pa}$, 
which is related to $h_{\perp}$ through (\ref{hw.pa.eq}) 
as a geometrical consequence of the non-stretching property of the flow, $\nabla_t|X|=0$, and the torsion-free property of the framing, $\nabla_tX=\nabla_xY$.

We thus emphasize that the formulation stated in Lemma \ref{lemma.cartan.eq} applies to {\it all} non-stretching curve flows $\gamma(t,x)$ in $\mb{HP}^n$, with the flow being determined by specifying $Y_{\perp}$ freely as a function of $t$ at each point $x$ along the curve. 

We now proceed with presenting
the bi-Hamiltonian operators encoded in the Cartan structure equations 
(\ref{uteq}), (\ref{wperpeq}), (\ref{hw.pa.eq}). 
Write 
\beq\label{hmhrel}
h^{\perp}:=\ad(e_x)h_{\perp}=\frac{1}{\sqrt{\chi}}\begin{pmatrix}2\hh_{\perp}&0&\tb{0}\\0&-2\hh_{\perp}&-\tb{h}_{\perp}\\\tb{0}&\ov{\tb{h}}^t&\tb{0}\end{pmatrix}:=(\hh^{\perp},\tb{h}^{\perp})\in\mk{h}_{\perp}
\eeq
which belongs to the same space as $u=(\u,\tbu)$, 
where 
\beq\label{h.pe.up}
\sqrt{\chi}\hh^{\perp}=2\hh_{\perp}\in\Q,\quad\sqrt{\chi}\tb{h}^{\perp}=-\tb{h}_{\perp}\in\mb{H}^{n-1}.
\eeq
Then the flow equation (\ref{uteq}) can be written in operator form \cite{AncoJGP}
\beq\label{ufloweq}
u_t = \Hop(\varpi^\perp) +h^\perp ,\quad
\varpi^\perp = \Jop(h^\perp) ,
\eeq
where
\beq
\Hop = \Kop|_{\h_\perp} ,\quad
\Jop= -\ad(\ehook{x})\inv \Kop|_{\m_\perp} \ad(\ehook{x})\inv
\label{HJops}
\eeq
are linear operators which act on $\h_\perp$-valued functions
and are invariant under $\equivH$,
as defined in terms of the operator
\beq\label{Kop}
\Kop :=
\Dx +[u,\cdot]_\perp -[u,\Dinvx[u,\cdot]_\parallel] .
\eeq

To display these operators explicitly, 
we first write out the flow on the scalar and vector variables $u=(\u,\tb{u})$,
yielding
\begin{align}
&\u_t=D_x\ww^{\perp}+C(\u,\ww^{\pa})+\ha C(\tb{u},\tb{w}^{\perp})+\hh^{\perp},
\label{f.on.u}\\
&\tb{u}_t=D_x\tb{w}^{\perp}-\ww^{\pa}\tb{u}+\tb{u}\tb{W}^{\pa}+\ww^{\perp}\tb{u}-\u\tb{w}^{\perp}+\tb{h}^{\perp},
\label{f.on.v.u}
\end{align}
where 
\begin{align}
&\ww^{\pa}=-D_x^{-1}\big(C(\u,\ww^{\perp})-\ha C(\tbu,\tbw^{\perp})\big),\qquad 
\tb{W}^{\pa}=D_x^{-1}\boldsymbol{\rm C}(\tbu,\tbw^{\perp}),
\label{w.pa}\\
&\ww^{\perp}=\chi\big(\frac{1}{4}D_x\hh^{\perp}+\frac{1}{4}C(\tbu,\tbh^{\perp})\big)+\hh_{\pa}\u,\qquad 
\tb{w}^{\perp}=\chi\big(D_x\tbh^{\perp}+\ha\hh^{\perp}\tbu+\u\tbh^{\perp}\big)+\hh_{\pa}\tbu,
\label{w.pe}\\
&\hh_{\pa}=-\chi D_x^{-1}\big(\ha A(\u,\hh^{\perp})-\ha A(\tbu,\tbh^{\perp})\big).
\label{hh.pa}
\end{align}
Next we introduce some  operator notations. 
For $\v\in\Q,\tb{v}\in\mb{H}^{n-1},\tb{V}\in\mk{u}(n-1,\mb{H})$, let 
\beq
\begin{aligned}
&C_{\u}\v:=C(\u,\v)\in\Q,\quad 
C_{\tb{u}}\tb{v}:=\ha C(\tb{u},\tb{v})\in\Q,
\\
&
\tb{C}_{\tb{u}}\tb{v}:=\tb{C}(\tb{u},\tb{v})\in\mk{u}(n-1,\mb{H}),\quad 
A_{\tb{u}}\tb{v}:=\ha A(\tb{u},\tb{v})\in\mb{R}
\end{aligned}
\eeq
denoting commutator and anti-commutator operators, 
and let
\beq
R_{\tb{u}}\v:=\v\tb{u}\in\mb{H}^{n-1},\quad 
L_{\u}\tb{v}:=\u\tb{v}\in\mb{H}^{n-1},\quad 
L_{\tb{u}}\tb{V}:=\tb{u}\tb{V}\in\mb{H}^{n-1}
\eeq
denoting right and left multiplication. 
Last it is convenient to scale the variables
\beq\label{scal.hame}
\hh^{\perp}\to\frac{1}{\chi}\hh^{\perp},\quad \tbh^{\perp}\to\frac{1}{\chi}\tbh^{\perp}
\eeq
in order to absorb all $\chi$ factors in (\ref{w.pe})--(\ref{hh.pa}).

\begin{theorem}\label{main.theorem}
The scaled flow equations (\ref{f.on.u}),(\ref{f.on.v.u}),(\ref{scal.hame}) 
for the quaternion variables $\u(t,x)\in\Q$ and $\tbu(t,x)\in\mb{H}^{n-1}$ 
have the operator form 
\beq\label{mail.th.eq}
\begin{pmatrix}\u_t\\\boldsymbol{\rm u}_t\end{pmatrix}=\mathcal{H}\begin{pmatrix}\ww^{\perp}\\\tbw^{\perp}\end{pmatrix}+\chi^{-1}\begin{pmatrix}\hh^{\perp}\\\tbh^{\perp}\end{pmatrix},\quad
\begin{pmatrix}\ww^{\perp}\\\tbw^{\perp}\end{pmatrix}=\mathcal{J}\begin{pmatrix}\hh^{\perp}\\\tbh^{\perp}\end{pmatrix},
\eeq
where
\beq\label{ham.op}
\mathcal{H}=\begin{pmatrix}D_x-C_{\u}D_x^{-1}C_{\u}&&&C_{\tbu}+C_{\u}D_x^{-1}C_{\tbu}\\\\
R_{\tbu}D_x^{-1}C_{\u}+R_{\tbu}&&&D_x-R_{\tbu}D_x^{-1}C_{\tbu}+L_{\tbu}D_x^{-1}\boldsymbol{\rm C}_{\tbu}-L_{\u}\end{pmatrix}
\eeq
and
\beq\label{sym.op}
\mathcal{J}=\begin{pmatrix}\frac{1}{4} D_x-\frac{1}{4}A_{\u}D_x^{-1}A_{\u}&&&\frac{1}{2}C_{\tbu}+\frac{1}{2} A_{\u}D_x^{-1}A_{\tbu}\\\\
\ha R_{\tbu}-\ha R_{\tbu}D_x^{-1}A_{\u}&&&D_x+L_{\u}+R_{\tbu}D_x^{-1}A_{\tbu}\end{pmatrix}
\eeq
are compatible Hamiltonian cosymplectic and symplectic operators on the $x$-jet space of $(\u,\tbu)$.
\end{theorem}

The proof of Theorem \ref{main.theorem} follows directly from  general results proven in \cite{AncoJGP} on the Hamiltonian structure of non-stretching curve flows in symmetric spaces, as applied to $\mb{HP}^n\simeq {\rm Sp}(n+1)/{\rm Sp}(1)\times {\rm Sp}(n)$. 
The work in \cite{AncoJGP} also develops the basic theory and properties of 
bi-Hamiltonian operators for Lie-algebra valued flow variables,
generalizing the standard treatment for scalar variables given in \cite{Olver}
(see also \cite{Dorfman}). 

To explain the definition and properties of Hamiltonian operators in the setting of quaternions, 
we start by defining variational derivatives on the $x$-jet space 
$J^{\infty}:=(x,\u,\tbu,\ov{\tbu},$ $\u_x,\tbu_x,\ov{\tbu}_x,\ldots)$ 
of the quaternionic flow variables $\u(t,x)$ and $\tbu(t,x)$. 
For any real-valued functional 
$\mk{H}=\int H(x,\u,\tbu,\ov{\tbu},\u_x,\tbu_x,\ov{\tbu}_x,\ldots)dx$,
its variational derivatives with respect to $\u$ and $\tbu$ 
are defined in terms of Frechet derivatives of $H(x,\u,\tbu,\ov{\tbu},\u_x,\tbu_x,\ov{\tbu}_x,\ldots)$ by 
\begin{align}
&\delta_{\hh^{\perp}}\mk{H}=\int {\rm pr}(\hh^{\perp}\cdot \p/\p\u)Hdx=:
\int{\rm Re}(\hh^{\perp}(\delta\mk{H}/\delta\u))dx={\rm Re}\int <\hh^{\perp},-\delta\mk{H}/\delta\u>dx,
\label{frechet.a}\\
&\delta_{\tbh^{\perp}}\mk{H}=\int {\rm pr}({\rm Re}(\tbh^{\perp}\cdot \p/\p\tbu))Hdx=:\int {\rm Re}(\tbh^{\perp}(\delta\mk{H}/\delta\tbu)^\t)=
{\rm Re}\int<\tbh^{\perp},\delta\mk{H}/\delta\ov{\tbu}>dx
\label{frechet.b}
\end{align}
modulo total $x$-derivatives, 
holding for all imaginary quaternion functions $\hh^{\perp}$ 
and all quaternionic vector functions $\tbh^{\perp}$, 
where $\hh^{\perp}\cdot \p/\p\u$ and 
${\rm Re}(\tbh^{\perp}\cdot \p/\p\tbu)=\ha(\tbh^{\perp}\cdot \p/\p\tbu+\ov{\tbh}^{\perp}\cdot \p/\p\ov{\tbu})$ 
are corresponding vector fields prolonged to $J^{\infty}$. 
(Here the dot denotes summation over quaternion components.) 
We note these definitions of $\delta \mk{H}/\delta\u$ and $\delta\mk{H}/\delta\tbu$ involve a reordering of products of quaternion variables
(as given by the middle equalities of (\ref{frechet.a}) and (\ref{frechet.b})),
for which we use the quaternionic multiplication and conjugation identities (\ref{iden.a})--(\ref{iden.f}). 

Now the property stated in Theorem \ref{main.theorem} that the operator $\Hop$ is cosymplectic means it defines an associated Poisson bracket 
\beq\label{poisson}
\{\mk{H}_1,\mk{H}_2\}:=
{Re}\int <\begin{pmatrix}-\delta \mk{H}_1/\delta\u\\\delta\mk{H}_1/\delta\ov{\tbu}\end{pmatrix}, \Hop 
\begin{pmatrix}-\delta \mk{H}_2/\delta\u\\\delta\mk{H}_2/\delta\ov{\tbu}\end{pmatrix}>dx
\eeq
which is skew-symmetric and obeys the Jacobi identity, for all  real-valued functionals $\mk{H}_1,\mk{H}_2$. 
A counterpart of the Poisson bracket is the symplectic $2$-form defined in terms of the operator $\Jop$ by 
\beq
\boldsymbol{\conx}({\rm X}_1,{\rm X}_2)_{\Jop}:={\rm Re}\int <
\begin{pmatrix}{\rm X}_1\u\\{\rm X}_1\tbu\end{pmatrix},\Jop 
\begin{pmatrix}{\rm X}_2\u\\{\rm X}_2\tbu\end{pmatrix}>dx
\eeq
where ${\rm X}_1,{\rm X}_2$ are vector fields ${\rm X}=\hh^{\perp}\cdot \p/\p\u+{\rm Re}(\tbh^{\perp}\cdot \p/\p\tbu)$ associated to pairs of an imaginary quaternion function $\hh^{\perp}$ and a quaternion vector function $\tbh^{\perp}$. 
The property stated in Theorem \ref{main.theorem} that $\Jop$ is symplectic corresponds to $\boldsymbol{\conx}$ being skew-symmetric and closed. 
In particular, closure means that
\beq\label{sym.cyclic}
\begin{aligned}
0 &=
{\rm pr}({\rm X}_1)\boldsymbol{\conx}({\rm X}_2,{\rm X}_3)+{\rm cyclic}\\
&=
{\rm Re}\int <\begin{pmatrix}\hh_2^{\perp}\\\tbh^{\perp}_2\end{pmatrix},
{\rm pr}(\hh^{\perp}_1\cdot \p/\p\u+{\rm Re}(\tbh^{\perp}_1\cdot \p/\p\tbu))\Jop 
\begin{pmatrix}\hh_3^{\perp}\\\tbh_3^{\perp}\end{pmatrix}>dx+{\rm cyclic}
\end{aligned}
\eeq
holds modulo total $x$-derivatives for all vector fields ${\rm X}_1,{\rm X}_2,{\rm X}_3$. 
Compatibility of these operators $\Hop$ and $\Jop$ is the statement that 
every linear combination $c_1\Hop+c_2\Jop^{-1}$ is a cosymplectic Hamiltonian operator,
or equivalently that $c_1\Hop^{-1}+c_2\Jop$ is a symplectic operator, 
where $\Hop^{-1}$ and $\Jop^{-1}$ denote formal inverse operators defined on the $x$-jet space $J^{\infty}$.

\section{Bi-Hamiltonian Hierarchies of Soliton Equations}

Composition of the compatible Hamiltonian operators (\ref{ham.op}) and (\ref{sym.op}) yields a recursion operator 
\beq\label{R.explicit}
\mathcal{R}:=\mathcal{H}\mathcal{J}=\begin{pmatrix}\mathcal{R}_{11}&\mathcal{R}_{12}\\\mathcal{R}_{21}&\mathcal{R}_{22}\end{pmatrix}
\eeq
given by 
\begin{align*}
\mathcal{R}_{11} &=
\tfrac{1}{4}D_x^2 +\ha C_{\tbu}R_{\tbu}
-\tfrac{1}{4}D_xA_{\u}D_x^{-1}A_{\u}-\tfrac{1}{4}C_{\u}D_x^{-1}C_{\u}D_x
+\ha C_{\u}D_x^{-1}C_{\tbu}R_{\tbu},
\\
\mathcal{R}_{12} &=
\ha D_xC_{\tbu} +C_{\tbu}D_x+C_{\tbu}L_{\u}
+\ha D_x A_{\u}D_x^{-1}A_{\tbu} +C_{\u}D_x^{-1}C_{\tbu}D_x
-\ha C_{\u}D_x^{-1}C_{\u}C_{\tbu}\\&\qquad
+C_{\u}D_x^{-1}C_{\tbu}L_{\u},
\\
\mathcal{R}_{21} &=
\ha D_xR_{\tbu} +\tfrac{1}{4}R_{\tbu}D_x -\ha L_{\u}R_{\tbu}
+\tfrac{1}{4}R_{\tbu}D_x^{-1}C_{\u}D_x -\ha D_xR_{\tbu}D_x^{-1}A_{\u}
-\tfrac{1}{4}R_{\tbu}A_{\u}D_x^{-1}A_{\u}
\\&\qquad
-\ha R_{\tbu}D_x^{-1}C_{\u}R_{\tbu}
+\ha L_{\tbu}D_x^{-1}\boldsymbol{\rm C}_{\tbu}R_{\tbu}
+\ha L_{\u}R_{\tbu}D_x^{-1}A_{\tbu},
\\
\mathcal{R}_{22} &=
D_x^2 +D_xL_{\u} -L_{\u}D_x -L_{\u}L_{\u} +\ha R_{\tbu}C_{\tbu} 
+D_xR_{\tbu}D_x^{-1}A_{\tbu} -R_{\tbu}D_x^{-1}C_{\tbu}D_x
+L_{\tbu}D_x^{-1}\boldsymbol{\rm C}_{\tbu}D_x
\\&\qquad
+\ha R_{\tbu}D_x^{-1}C_{\u}C_{\tbu} +\ha R_{\tbu}A_{\u}D_x^{-1}A_{\tbu}
-R_{\tbu}D_x^{-1}C_{\tbu}L_{\u}
+L_{\tbu}D_x^{-1}\boldsymbol{\rm C}_{\tbu}L_{\u}
-L_{u}R_{\tbu}D_x^{-1}A_{\tbu} 
.
\end{align*}
Each of these operators $\mathcal{H},\mathcal{J},\mathcal{R}$ displays obvious symmetry invariance under translations in $x$.
As a consequence, from general results due to Magri \cite{Magri1978,Magri1980},
the recursion operator will generate a hierarchy of 
commuting Hamiltonian vector fields with respect to the Poisson bracket, 
starting from the evolutionary form of the $x$-translation vector field $\p/\p x$. 
Moreover, the adjoint recursion operator $\mathcal{R}^*$
will generate an involutive hierarchy of variational covector fields, 
arising from the canonical pairing provided by the symplectic $2$-form.
This leads to the following results \cite{AncoJGP}.

\begin{theorem}\label{th.5}
The pairs of quaternionic scalar-vector functions $\hh_{(l)}^{\perp}\in\Q,\tbh_{(l)}^{\perp}\in\mb{H}^{n-1}$ given by 
\beq\label{trivial.sym}
\begin{pmatrix}\hh_{(l)}^{\perp}\\\tbh_{(l)}^{\perp}\end{pmatrix}:=\mathcal{R}^l\begin{pmatrix}\u_x\\\tbu_x\end{pmatrix},\qquad l=0,1,2,\ldots, 
\eeq
yield a commuting  hierarchy of Hamiltonian vector fields $\hh_{(l)}^{\perp}\cdot\p/\p\u+{\rm Re}(\tbh_{(l)}^{\perp}\cdot\p/\p\tbu)$.
In particular, there exists corresponding Hamiltonian functionals $\mk{H}^{(l)}$ such that   
$$ \delta_{(\hh^{\perp}_{(l)},\tbh_{(l)}^{\perp})}\mk{G}=\{\mk{G},\mk{H}^{(l)}\}_{\Hop} $$ 
for all functionals $\mk{G}$ on $J^{\infty}$. 
Explicit expressions for the Hamiltonians 
$H^{(l)}\in\mb{R}$, $l=0,1,2,\ldots$, 
are given by 
\beq\label{Hl.a}
\begin{aligned}
H^{(l)}: &= 
\frac{1}{1+2l}D_x^{-1}{\rm Re}(<\u,\hh^{\perp}_{(l)}>+<\tbu,\tbh_{(l)}^{\perp}>)\\&
=\frac{1}{2+4l}D_x^{-1}(-A(\u,\hh^{\perp}_{(l)})+A(\tbu,\tbh_{(l)}^{\perp}))=\frac{1}{1+2l}\hh_{\pa}^{(l)}
\end{aligned}
\eeq
whose variational derivatives 
$$\ww^{\perp}_{(l)}:=\delta H^{(l)}/\delta \ov{\u}=-\delta H^{(l)}/\delta \u\in\Q,\quad \tbw^{\perp}_{(l)}:=\delta H^{(l)}/\delta \ov{\tbu}\in\mb{H}^{n-1}$$
yield an associated hierarchy of involutive covector fields 
$\ww^{\perp}_{(l)}\cdot d\u+{\rm Re}(\tbw^{\perp}_{(l)}\cdot d\tbu)$
given by 
\beq\label{Wl.b}
\begin{pmatrix}\ww_{(l)}^{\perp}\\\tbw_{(l)}^{\perp}\end{pmatrix}:=\mathcal{R}^{*l}\begin{pmatrix}\u\\\tbu\end{pmatrix},\qquad l=0,1,2,\ldots.
\eeq
These variational covector fields are dual to the Hamiltonian vector fields 
via the pairing
\beq
(\ww^{\perp}_{(l)}\cdot d\u+{\rm Re}(\tbw^{\perp}_{(l)}\cdot d\tbu))\rfloor {\rm X}=
\boldsymbol{\conx}({\rm X},\hh^{\perp}_{(l)}\cdot\p/\p\u+{\rm Re}(\tbh^{\perp}_{(l)}\cdot \p/\p\tbu))
\eeq
holding for all vector fields ${\rm X}$ in evolutionary form on $J^{\infty}$.
\end{theorem}

Note both hierarchies (\ref{trivial.sym}) and (\ref{Wl.b}) 
possess the mKdV scaling symmetry 
$x\to\lambda x$, $(\u,\tbu)\to(\lambda^{-1}\u,\lambda^{-1}\tbu)$, 
with (\ref{trivial.sym}) and (\ref{Hl.a}) having the scaling weight $2+2l$ 
and (\ref{Wl.b}) having the scaling weight $1+2l$.

We can now state our main result.

\begin{theorem}\label{th.5.2}
The flow equations (\ref{mail.th.eq}) on the imaginary scalar quaternion variable $\u(t,x)$ and the $n-1$-component quaternion vector variable $\tbu(t,x)$ 
yield a hierarchy of bi-Hamiltonian evolution equations 
\beq\label{hier.ar}
\begin{pmatrix}\u_t\\\tbu_t   \end{pmatrix}=
\begin{pmatrix}\hh_{(l)}^{\perp}+\chi^{-1}\hh_{(l-1)}^{\perp}\\
\tbh_{(l)}^{\perp}+\chi^{-1}\tbh_{(l-1)}^{\perp}\end{pmatrix},\quad l=1,2,\ldots,
\eeq
called the $(+l)$-flow, with the Hamiltonian structure 
\begin{align}
\begin{pmatrix}\hh_{(l)}^{\perp}\\\tbh_{(l)}^{\perp}\end{pmatrix} 
&=
\mathcal{H}\begin{pmatrix}-\delta H^{(l)}/\delta \u\\\delta H^{(l)}/\delta \ov{\tbu}\end{pmatrix},\quad l=0,1,2,\ldots,
\label{hier.ar.1}\\
&=
\mathcal{E}\begin{pmatrix}-\delta H^{(l-1)}/\delta \u\\\delta H^{(l-1)}/\delta \ov{\tbu}\end{pmatrix},\quad l=1,2,\ldots,
\label{hier.ar.2}
\end{align}
where $\mathcal{E}:=\mathcal{R}\mathcal{H}$.
Each of these multi-component quaternionic evolution equations (\ref{hier.ar}) 
is invariant under the group 
$H^*_{\pa}\simeq \Ad({\rm U}(1,\mb{H})\times {\rm U}(n-1,\mb{H}))$ 
given by rigid ($x$-independent) transformations (\ref{gauge.tr.1}) 
on the pair of scalar-vector variables $(\u,\tbu)$.
\end{theorem}
 
\subsection{mKdV flow}

The $+1$ flow in the hierarchy (\ref{hier.ar}) comes from 
\beq\label{tri.sym.x}
\begin{pmatrix}\hh_{(0)}^{\perp}\\\tbh_{(0)}^{\perp}\end{pmatrix}=\begin{pmatrix}\u_x\\\tbu_x\end{pmatrix}
\eeq
producing a coupled system of quaternionic scalar-vector mKdV equations 
\begin{align}
& \u_t-\chi^{-1} \u_x=
\tfrac{1}{4}\u_{3x}-\tfrac{3}{2}\u^2\u_x+\tfrac{3}{4}C(\u,C(\tb{u},\tb{u}_x))+
\tfrac{3}{4}C(\tb{u},\tb{u}_{2x})=\hh_{(1)}^{\perp},
\label{mkdv.eq}\\
& \tb{u}_t-\chi^{-1}\tbu_x=
\tb{u}_{3x}+\tfrac{3}{2}(|\tb{u}|^2-\u^2+\u_x)\tb{u}_x
+\tfrac{3}{4}(2\u|\tb{u}|^2-A(\u,\u_x)+\u_{2x}-C(\tb{u},\tb{u}_x))\tb{u}=\tbh_{(1)}^{\perp},
\label{mkdv.eq.v}
\end{align}
where $C(\cdot,\cdot)$ and $A(\cdot,\cdot)$ respectively denote 
the commutators and anticommutators defined in (\ref{not.com.anti})--(\ref{not.com.mat}). 

This is an integrable system in the following sense. 
We first note that the convective terms $\u_x$ and $\tbu_x$ in both equations (\ref{mkdv.eq})--(\ref{mkdv.eq.v}) can be removed by a Galilean transformation $t\to t,x\to x+\chi^{-1} t$. 
The resulting evolution equations then have the explicit bi-Hamiltonian structure  
given by the $l=1$ case of (\ref{hier.ar.1})--(\ref{hier.ar.2}):
\beq\label{mkd.biham}
\begin{aligned}
\begin{pmatrix}\u_t\\\tbu_t\end{pmatrix} 
&=
\mathcal{H}\begin{pmatrix}\delta H^{(1)}/\delta \u\\\delta H^{(1)}/\delta \tbu\end{pmatrix}=\mathcal{E}\begin{pmatrix}\delta H^{(0)}/\delta \u\\\delta H^{(0)}/\delta \tbu\end{pmatrix} 
\\
&=
\begin{pmatrix}
\tfrac{1}{4}\u_{3x}-\tfrac{3}{2}\u^2\u_x+\tfrac{3}{4}C(\u,C(\tb{u},\tb{u}_x))
+\tfrac{3}{4}C(\tb{u},\tb{u}_{2x})\\
\tb{u}_{3x}+\tfrac{3}{2}(|\tb{u}|^2-\u^2+\u_x)\tb{u}_x
+\tfrac{3}{4}(2 \u|\tb{u}|^2-A(\u,\u_x)-C(\tb{u},\tb{u}_x)+\u_{2x})\tb{u}
\end{pmatrix}
\end{aligned}
\eeq
where 
\beq\label{hamilton.func}
H^{(0)}=-\ha \u^2+|\tbu|^2\quad\textrm{and}\quad 
H^{(1)}=\tfrac{1}{8}\u_x^2-\ha|\tbu_x|^2 -\tfrac{1}{8}A(\u,C(\tbu,\tbu_x))
+\tfrac{1}{8}(\u^2-|\tbu|^2)^2
\eeq
are the Hamiltonians. 

In addition to its explicit symmetry with respect to space translations  $x\to x+\epsilon$ and mKdV scalings 
$x\to\lambda x$, $t\to\lambda^3 t$, $\u\to \lambda^{-1}\u$, $\tbu\to\lambda^{-1}\tbu$, 
the quaternion scalar-vector mKdV system (\ref{mkd.biham})  
possesses higher symmetries given by each higher-order flow in the hierarchy, i.e. 
${\rm X}=\hh^{\perp}_{(l)}\cdot\p/\p\u+{\rm Re}(\tbh_{(l)}^{\perp}\cdot \p/\p\tbu)$ 
generates an infinitesimal symmetry of the coupled system (\ref{mkd.biham}) for all $l=2,3,\ldots$. 
Moreover, 
each Hamiltonian in the hierarchy yields a conservation law for this system (\ref{mkd.biham}), i.e. 
$\frac{d}{dt}\int_{-\infty}^{\infty}H^{(l)}dx=0$ for $l=0,1,2,\ldots$ 
holds for all solutions $\u(t,x),\tbu(t,x)$ that have sufficiently fast decay as $x\to \pm \infty$.

\subsection{SG flow}
Apart from the $+1,+2,\ldots$ flows in the mKdV hierarchy (\ref{hier.ar}), 
the recursion operator $\mathcal{R}=\mathcal{H}\mathcal{J}$ also yields a flow defined by 
\beq\label{SG.om.0}
0=\begin{pmatrix}\ww^{\perp}\\\tbw^{\perp}\end{pmatrix}=\mathcal{J}\begin{pmatrix}\hh^{\perp}\\\tbh^{\perp}\end{pmatrix}.
\eeq 
This will be called the $-1$ flow \cite{AncoJGP}. 
The resulting flow equations (\ref{mail.th.eq}) have the form 
\beq\label{sg.10}
\begin{pmatrix}\u_t\\\tbu_t\end{pmatrix}=\chi^{-1}\begin{pmatrix}\hh^{\perp}\\\tbu^{\perp}\end{pmatrix}
\eeq
with 
\beq\label{sg.11}
D_x\hh^{\perp}=-C(\tbu,\tbh^{\perp})-4\hh_{\pa}\u,\quad D_x\tbh^{\perp}=-\ha \hh^{\perp}\tbu-\u\tbh^{\perp}-\hh_{\pa}\tbu
\eeq
and 
\beq\label{sg.12}
D_x\hh_{\pa}=-\ha A(\u,\hh^{\perp})-\ha A(\tbu,\tbh^{\perp}).
\eeq
These equations (\ref{sg.11})--(\ref{sg.12}) possess the conservation law 
\beq\label{sg.13}
D_x(\hh_{\pa}^2+\tfrac{1}{4}|\hh^{\perp}|^2+|\tbh^{\perp}|^2)=0.
\eeq
Hence, 
after a conformal scaling of $t$, which induces a corresponding scaling of the variables $\hh_{\pa},\hh^{\perp},\tbh^{\perp}$ by a function of $t$, 
we get 
\beq\label{sg.14}
\hh_{\pa}^2+\tfrac{1}{4}|\hh^{\perp}|^2+|\tbh^{\perp}|^2=c={\rm const}.
\eeq
yielding the relation 
\beq\label{sg.15}
\hh_{\pa}=\pm\sqrt{c-\tfrac{1}{4}|\hh^{\perp}|^2-|\tbh^{\perp}|^2}.
\eeq
Substitution of (\ref{sg.15}) and (\ref{sg.10}) into (\ref{sg.11}) then gives 
a coupled hyperbolic system of quaternionic scalar-vector SG equations 
\begin{align}
&\u_{tx}=\mp4\sqrt{(1-\tfrac{1}{4}|\u_t|^2-|\tb{u}_t|^2)}\ \u-C(\tb{u},\tb{u}_t),
\label{SG.eq}\\
&\tb{u}_{tx}=\mp\sqrt{(1-\tfrac{1}{4}|\u_t|^2-|\tb{u}_t|^2)}\ \tb{u}-\ha\u_t\tb{u}-\u\tb{u}_t,
\label{SG.eq.v}
\end{align}
in which we have put $c=\chi^2$ without loss of generality
and scaled out a factor $\chi$ by means of the transformation 
$x\to \chi^{-1} x$. 
Here $C(\cdot,\cdot)$ denotes the commutator defined in (\ref{not.com.vec}). 

This system (\ref{SG.eq})--(\ref{SG.eq.v}) has explicit symmetry under 
separate time translations $t\to t+\epsilon$ and space translations $x\to x+\epsilon$ 
as well as scalings 
$t\to\lambda^{-1}t$, $\u\to\lambda^{-1}\u$, $\tbu\to\lambda^{-1}\tbu$.  
It also possesses higher symmetries 
${\rm X}=\hh^{\perp}_{(l)}\cdot \p/\p\u+{\rm Re}(\tbh^{\perp}_{(l)}\cdot\p/\p\tbu)$ for $l=2,3,\ldots$, 
given by each higher-order flow (\ref{trivial.sym}) 
in the quaternionic mKdV hierarchy (\ref{trivial.sym}), 
along with conservation laws  
$\frac{d}{dt}\int_{-\infty}^{\infty}H^{(l)}dx=0$ for $l=0,1,2,\ldots$, 
given by the corresponding Hamiltonians (\ref{Hl.a}) in the same hierarchy, as verified through (\ref{sg.10})--(\ref{sg.12}).

\subsection{Reductions and Soliton equations in $\mb{HP}^1$}

We now make some remarks on reductions of the quaternionic mKdV system (\ref{mkd.biham}) and quaternionic SG system (\ref{SG.eq}).

Firstly, if we consider a reduction to a vector system by putting $\u=0$, 
then the quaternion vector inner product terms 
$C(\tbu,\tbu_{2x})$ and $C(\tbu,\tbu_t)$ in the respective equations 
(\ref{mkd.biham}) and (\ref{SG.eq}) for the quaternion scalar variable $\u$ 
need to vanish identically. 
Such algebraic constraints cannot be satisfied unless we restrict the vector variable $\tbu$ to be  commutative by taking its components to belong to a real or complex subalgebra of $\mb{H}$. 
Therefore, 
both the quaternion mKdV system (\ref{mkd.biham}) and quaternion SG system (\ref{SG.eq}) have {\it no} consistent non-commutative vector reduction.
The underlying reason why the quaternion vector variable $\tbu$ is unavoidably coupled to the quaternion scalar variable $\u$ can be understood from 
the nontrivial Lie bracket relation 
$[(0,\tbh_{1\perp}),(0,\tbh_{2\perp})]_{\pa}=\big(\ha C(\tbh_{1\perp},\tbh_{2\perp}),\mathbf{0}\big)$ for the space $\mk{h}_{\perp}$ 
in which $\tbu$ lies (cf. (\ref{inclu.11}) and (\ref{mat.con})).

In contrast, we can get a consistent scalar  reduction by putting $\tbu=0$ in the systems (\ref{mkd.biham}) and (\ref{SG.eq}). 
This reduction has a natural geometric meaning if non-stretching curve flows  are considered in a submanifold $\mb{HP}^1\subset \mb{HP}^n$ or in $\mb{HP}^1$ itself when $n=1$.
 
Thereby we obtain a scalar non-commutative mKdV equation 
\beq\label{red.mkdv}
\u_t=\tfrac{1}{4}\u_{3x}-\tfrac{3}{2}\u^2\u_x
\eeq
and a scalar non-commutative SG equation 
\beq\label{red.sg}
\u_{tx}= 4\u\sqrt{1+\tfrac{1}{4}\u_t^2}
\eeq
in which $\u(t,x)$ is an imaginary quaternion variable. 
We note the underlying Hamiltonian structure 
in this case is given by Theorems \ref{th.5} and \ref{th.5.2} with $\tbu=0$ and $\tbh^{\perp}=0$, where the Hamiltonian operators take form 
\beq\label{red.ham.sym}
\mathcal{H}=D_x-C_{\u}D_x^{-1}C_{\u},\quad 
\mathcal{J}=\tfrac{1}{4} D_x-\tfrac{1}{4}A_{\u}D_x^{-1}A_{\u}
\eeq
(which are the first diagonal entries of (\ref{ham.op}) and (\ref{sym.op}), 
respectively).

\section{Geometric Curve Flows}

The bi-Hamiltonian flows given in Theorems~\ref{th.5} and~\ref{th.5.2} 
have a direct geometrical formulation in terms of the 
${\rm U}(1,\mb{H})\times{\rm U}(n,\mb{H})$-parallel frame variables 
(\ref{pull.not.1}), (\ref{pull.not.2}), (\ref{mat.con}) 
describing a non-stretching flow of a curve $\gamma$ in $M=\mb{HP}^n$. 
To begin, 
we recall that the quaternionic components of the connection matrix $\conx_x=(\u,\tbu)=u$ describe covariants of $\gamma$ relative to the equivalence group of the framing. 
This means that $u$ is invariantly determined by $\gamma$ up to the action of the group 
\beq\label{act.gr}
H_{\pa}^*\simeq {\rm Ad({\rm U}(1,\mb{H})\times{\rm U}(n-1,\mb{H}))}
\eeq
as given by the rigid ($x$-independent) transformations (\ref{gauge.tr.1}). 
The evolution of $u$ under the bi-Hamiltonian flows (\ref{hier.ar})--(\ref{hier.ar.2}) is given by 
\beq\label{ac.gr.1}
u_t=h^{\perp}_{(l+1)}+\chi^{-1} h^{\perp}_{(l)}
\eeq
with 
\beq\label{act.gr.2}
h^{\perp}_{(l)}=\mathcal{R}^lu_x
\eeq
for all $l=0,1,2,\ldots$, where $\mathcal{R}$ is the recursion operator defined in terms of $u$ through (\ref{HJops}) and (\ref{Kop}).  
Each evolution equation (\ref{ac.gr.1}) is invariant with respect to the transformations (\ref{gauge.tr.1})
and belongs to the general class of flows in which
\beq\label{equi.flow}
\begin{aligned}
h^{\perp}(x,u,u_x,u_{xx},\ldots)={\rm Ad}(a^{-1})h^{\perp}(x,{\rm Ad}(a)u,{\rm Ad}(a)u_x,{\rm Ad}(a)u_{xx},\ldots), \\ 
a\in{\rm U}(1,\mb{H})\times{\rm U}(n-1,\mb{H})
\end{aligned}
\eeq
is an equivariant function of the invariant arclength $x$ of $\gamma$ 
and the (differential) covariants $u,u_x,u_{xx},\ldots$ of $\gamma$ relative to the group (\ref{act.gr}).
 
Any flow specified by the ${\rm U}(1,\mb{H})\times{\rm U}(n-1,\mb{H})$-equivariant class (\ref{equi.flow}) yields a non-stretching curve flow in $M=\mb{HP}^n$ via the geometric relations 
\beq\label{flow.from.flow}
h_{\perp}=\chi^{-1}\ad(e_x)^{-1}h^{\perp}=e\rfloor Y_{\perp}, \quad
h_{\pa}=-D_x^{-1}[u,h_{\perp}]_{\pa}=e\rfloor Y_{\pa}
\eeq
taking into account (\ref{hmhrel}) and (\ref{scal.hame}),
where $Y_{\perp}$ and $Y_{\pa}$ are the normal and tangential projections of $Y=\gamma_t$ relative to the tangent vector $X=\gamma_x$ along $\gamma$. 
In particular, the evolution vector of the curve is given by 
\beq\label{mat.cal.Y}
Y=-\<e^*,h_{\perp}+h_{\pa}\>=\chi^{-1}\<e^*,\mathcal{Y}(h^{\perp})\>
\eeq
in terms of the operator 
\beq\label{mat.cal.Y.2}
\mathcal{Y}:=D_x^{-1}[u,\ad(e_x)^{-1}\cdot]_{\pa}-\ad(e_x)^{-1}
\eeq
where $e^*$ is the linear frame dual to the linear coframe $e$ along $\gamma$, 
with $e_x=e\rfloor X$. 
In this relation (\ref{mat.cal.Y}), 
$e_x$ is preserved under the action of the equivalence group (\ref{act.gr}), 
while up to equivalence, 
both $e^*$ and $e$ are determined by $\conx_x$ through the transport equation (\ref{frenet.eq}) along $\gamma$. 
Hence we obtain the following geometric characterization of the class of flows (\ref{equi.flow}).  

\begin{proposition}
Every ${\rm U}(1,\mb{H})\times{\rm U}(n-1,\mb{H})$-equivariant flow (\ref{equi.flow}) 
determines a non-stretching curve flow $\gamma(t,x)$ in $\mb{HP}^n$ 
satisfying a ${\rm U}(n+1,\mb{H})$-invariant evolution equation 
\beq\label{g.geom.d}
\gamma_t=Y(x,\gamma_x,\nabla_x\gamma_x,\nabla^2_x\gamma_x,\ldots)
\eeq
through (\ref{mat.cal.Y})--(\ref{mat.cal.Y.2}),
where $x$ is the ${\rm U}(n+1,\mb{H})$-invariant arclength along the curve. 
\end{proposition} 

Invariance of an evolution equation (\ref{g.geom.d}) can be shown to imply that the function $Y$ is constructed using only the metric $g$  given by (\ref{metr.conn}) and the tensor $\ad_x^2$  defined as follows:
\beq\label{li.map.g}
e\rfloor\ad_x^2(X)Z:=\ad(e_X)^2e_Z
\eeq
for all $X,Z$ in $T_x M$. 
Note both $g$ and $\ad_x^2$ are gauge-invariant under changes of frame (\ref{gauge.tr}) and thus are well-defined as 
geometrical structures on the manifold $M=\mb{HP}^n$.

We can now formulate the bi-Hamiltonian flows from theorem \ref{th.5} and \ref{th.5.2}  in strictly geometrical terms. 

\begin{theorem}
The hierarchy of bi-Hamiltonian quaternionic flows (\ref{hier.ar})--(\ref{hier.ar.2}) correspond to  non-stretching geometric curve flows in $M=\mb{HP}^n$ given by evolution equations of the form 
$$ \gamma_t=Y_{(l)}(\gamma_x,\nabla_x\gamma_x,\nabla_x^2\gamma_x,\ldots), 
\quad |\gamma_x|=1, \quad (l=1,2,\ldots) $$
where each equation is invariant with respect to the isometry group 
${\rm U}(n+1,\mb{H})$ of $M=\mb{HP}^n$ 
and preserves the invariant arclength $x$. 
\end{theorem}

To write out these geometric curve flow equations explicitly, 
it is useful to introduce the linear map 
\beq\label{li.cal.X.h}
\mathcal{X}_{\gamma}:=-\ad_x^2(\gamma_x)
\eeq
which is determined by the curve $\gamma$. 
In a ${\rm U}(1,\mb{H})\times{\rm U}(n-1,\mb{H})$-parallel frame, 
this map corresponds to $-\ad(e_x)^2$ under which the vector space 
$\mk{m}=\mk{u}(n+1,\mb{H})/(\mk{u}(1,\mb{H})\oplus\mk{u}(n,\mb{H}))$ 
decomposes into a direct sum of eigenspaces 
$\mk{m}_{\pa}=\mb{R}$, $\mk{m}^s_{\perp}=\Q$, $\mk{m}_{\perp}^v=\mb{H}^{n-1}$ 
with respective eigenvalues $0$, $4/\chi$, $1/\chi$, 
where $\mk{m}_{\pa}$ is the centralizer space of $e_x$ in $\mk{m}$, 
and $\mk{m}_{\perp}=\mk{m}^s_{\perp}\oplus\mk{m}_{\perp}^v$ is the perp space of $\mk{m}_{\pa}$. 
Since the linear coframe $e$ provides an identification between $\mk{m}$ and $T_x M$, 
there is a corresponding decomposition of the tangent spaces $T_{\gamma}M$ along $\gamma$ given by 
\begin{gather*}
T_{\gamma}M=(T_{\gamma}M)_{\pa}\oplus(T_{\gamma}M)_{\perp},\quad 
(T_{\gamma}M)_{\pa}={\rm span}(\gamma_x),\\
(T_{\gamma}M)_{\perp}={\rm span}(\gamma_x)^{\perp}=(T_{\gamma}M)^s_{\perp}\oplus(T_{\gamma}M)^v_{\perp}
\end{gather*}
where $(T_{\gamma}M)_{\pa}\simeq \mb{R}$, $(T_{\gamma}M)^s_{\perp}\simeq \Q$, $(T_{\gamma}M)^v_{\perp}\simeq \mb{H}^{n-1}$ are the eigenspaces of $\mathcal{X}_{\gamma}$ with eigenvalues $0$, $4/\chi$, $1/\chi$. 

Hereafter, we will write the tangent vector of $\gamma$ as $T=\gamma_x\in(T_{\gamma}M)_{\pa}$ and the principal normal vector along $\gamma$ as $N=\nabla_x\gamma_x\in(T_{\gamma}M)_{\perp}$; 
$N^s$ and $N^v$ will denote the projections of $N$ into $(T_{\gamma}M)_{\pa}^s$ and $(T_{\gamma}M)_{\perp}^v$ respectively.

\subsection{mKdV Curve Flow}

From the flow (\ref{tri.sym.x}), 
after undoing the scaling (\ref{scal.hame}),
we have $\sqrt{\chi}\hh_{\perp}=\ha \u_x$, $\sqrt{\chi}\tbh_{\perp}=-\tbu_x$ 
through (\ref{h.pe.up}), 
and $\hh_{\pa}=-\ha \u^2+\ha|\tbu|^2$ through (\ref{hh.pa}).
The frame variables (\ref{h.pa.pe}) and (\ref{hh.pa.pe}) thereby yield 
\begin{gather}
(e_t)^s_{\perp}=\frac{1}{\sqrt{\chi}}\begin{pmatrix}0&\ha\u_x&\tb{0}\\\ha\u_x&0&\tb{0}\\\tb{0}&\tb{0}&\tb{0}\end{pmatrix},\quad 
(e_t)^v_{\perp}=\frac{1}{\sqrt{\chi}}\begin{pmatrix}0&0&-\tb{u}_x\\0&0&\tb{0}\\\ov{\tb{u}}_x^\t&\tb{0}&\tb{0}\end{pmatrix},
\label{e.t.sv.a}\\
(e_t)_{\pa}=(-\ha \u^2+\ha|\tbu|^2)e_x.
\label{e.t.sv.b}
\end{gather}
These quaternionic matrices can be expressed in terms of the geometrical vectors 
$T$, $N$, $\nabla_xN=:N'$ as follows. 
We first note from (\ref{ad.om}) and (\ref{ad.om.1}):
\beq\label{eN.sv.pa}
(e_{N})^s_{\perp}=\frac{1}{\sqrt{\chi}}\begin{pmatrix}0&2\u&\tb{0}\\2\u&0&\tb{0}\\\tb{0}&\tb{0}&\tb{0}\end{pmatrix}=e\rfloor N^s,\quad 
(e_N)^v_{\perp}=\frac{1}{\sqrt{\chi}}\begin{pmatrix}0&0&-\tb{u}\\0&0&\tb{0}\\\ov{\tb{u}}^\t&\tb{0}&\tb{0}\end{pmatrix}=e\rfloor N^v,\quad (e_N)_{\pa}=0,
\eeq
and 
\begin{gather}
(e_{N'})_{\pa}=(4\u^2-|\tbu|^2)e_x,\quad 
(e_{N'})^s_{\perp}=
\frac{1}{\sqrt{\chi}}
\begin{pmatrix}0&2\u_x&\tb{0}\\2\u_x&0&\tb{0}\\\tb{0}&\tb{0}&\tb{0}\end{pmatrix}
=e\rfloor (\nabla_xN)^s_{\perp},
\label{enN.v.pa}\\
(e_{N'})^v_{\perp}=
\frac{1}{\sqrt{\chi}}
\begin{pmatrix}0&0&-\tb{u}_x-3\u\tbu\\0&0&\tb{0}\\\ov{\tb{u}}_x^\t-3\ov{\tbu}^{\t}\u&\tb{0}&\tb{0}\end{pmatrix}
=e\rfloor (\nabla_xN)_{\perp}^v .
\label{enN.v.pe}
\end{gather}
In addition, we note:
\beq\label{ad.eN.ex}
(\ad(e_N)^2e_x)_{\perp}=
\frac{1}{\sqrt{\chi}^3}
\begin{pmatrix}0&0&-6\u\tbu\\0&0&\tb{0}\\-6\ov{\tbu}^{\t}\u&\tb{0}&\tb{0}\end{pmatrix}
=e\rfloor(\ad_x^2(N)T)_{\perp}^v .
\eeq
Since $|\tbu|^2=g(N^v,N^v)=|N^v|^2$ and $-4\u^2=g(N^s,N^s)=|N^s|^2$, then from (\ref{e.t.sv.b}) and (\ref{nn.g}) we find 
\beq\label{at.pa.dd}
(e_t)_{\pa}=(\tfrac{1}{8}|N^s|^2+\ha|N^v|^2)e_x=e\rfloor\big((\tfrac{1}{8}|N^s|^2+\ha|N^v|^2)T\big).
\eeq
Next, comparing (\ref{e.t.sv.a}) to (\ref{enN.v.pa}), (\ref{enN.v.pe}), (\ref{ad.eN.ex}), 
we obtain
\beq\label{et.s.pe.e}
(e_t)_{\perp}^s=\tfrac{1}{4}(e_{N'})_{\perp}^s=e\rfloor(\tfrac{1}{4}\nabla_xN)_{\perp}^s
\eeq
and
\beq\label{et.v.pe.f}
(e_t)_{\perp}^v=(e_{N'})^v_{\perp}-\ha\chi(\ad(e_N)^2e_x)_{\perp}^v=e\rfloor\big((\nabla_xN)_{\perp}^v-\ha\chi(\ad_x^2(N)T)_{\perp}^v\big).
\eeq

Thus, from (\ref{at.pa.dd}), (\ref{et.s.pe.e}), (\ref{et.v.pe.f}) 
combined with $e_t=e\rfloor \gamma_t$, 
we derive the following geometrical evolution equation 
\begin{equation}\label{gamma.t.g}
\gamma_t=\tfrac{1}{4}N'{}^s_{\perp}+N'{}^v_{\perp}-\ha\chi(\ad_x^2(N)T)_{\perp}^v+(\tfrac{1}{8}|N^s|^2+\ha|N^v|^2)T
\end{equation}
in terms of $T=\gamma_x$, $N=\nabla_xT$, $N'=\nabla_xN$. 
We can write this evolution equation in an equivalent form without the $s$ and $v$ projections, by relating (\ref{e.t.sv.a}) to $\ad(e_x)^{-2}$ applied to 
(\ref{eN.sv.pa}), (\ref{enN.v.pa}), (\ref{enN.v.pe}). 
This leads to the geometrical evolution equation 
\beq\label{gamma.t.h}
\chi\gamma_t=\nabla_x(\mathcal{X}_\gamma^{-1}\nabla_x\gamma_x)-\big(\ad_x^2(\mathcal{X}_\gamma^{-1}\nabla_x\gamma_x)\gamma_x\big)_{\perp}-3\big(\ad_x^2(\mathcal{X}_\gamma^{-1}\nabla_x\gamma_x)\gamma_x\big)_{\pa},\quad |\gamma_x|=1,
\eeq
or equivalently 
\beq\label{starstar}
\gamma_t=\mathcal{X}_\gamma^{-1}\big(\chi^{-1}\nabla_x^2\gamma_x-\ha\ad_x^2(\nabla_x\gamma_x)\gamma_x\big)_{\perp}-\ha\chi^{-1} g(\mathcal{X}_\gamma^{-1}\nabla_x\gamma_x,\nabla_x\gamma_x)\gamma_x,\quad |\gamma_x|=1,
\eeq
called the {\it non-stretching mKdV map} on $M=\mb{HP}^n$. 
Each equation (\ref{gamma.t.g}), (\ref{gamma.t.h}), (\ref{starstar}) 
is invariant under the isometry group ${\rm U}(n+1,\mb{H})$ of $M=\mb{HP}^n$.

\subsection{SG Curve Flow}

The flow given by (\ref{SG.om.0}) has $\ww^{\perp}=\tbw^{\perp}=0$ 
which implies $\ww^{\pa}=\tb{W}^{\pa}=0$ from (\ref{uteq}). 
Hence the frame variables (\ref{var.pa.f}) and (\ref{var.pe.f}) yield 
\begin{equation}\label{SG.cu.a}
(\conx_t)_{\perp}=(\conx_t)_{\pa}=0,
\end{equation}
so thus the connection matrix in the flow direction vanishes. 
This can be expressed geometrically in terms of the tangent vector $T=\gamma_x$ as follows. 
By applying $\ad(e_x)$ to $\conx_t=0$, 
we get 
\begin{equation}\label{SG.cu.b}
0=\ad(e_x)\conx_t=-[\conx_t,e_x]=e\rfloor(\nabla_tT)
\end{equation}
through using $D_te_x=0$. 
As a result, we obtain the geometrical evolution equation 
\begin{equation}\label{SG.cu.c}
0=\nabla_tT
\end{equation}
or equivalently 
\begin{equation}\label{SG.cu.d}
0=\nabla_t\gamma_x=\nabla_x\gamma_t,\quad |\gamma_x|=1,
\end{equation}
which is called the {\it non-stretching wave map} on $M=\mb{HP}^n$.  
In addition to satisfying the non-stretching property $\nabla_t|\gamma_x|=0$, 
this equation (\ref{SG.cu.d}) possesses the conservation law 
$\nabla_x|\gamma_t|=0$, 
corresponding to (\ref{sg.13}). 
Thus, up to a conformal scaling of $t$, the evolution given by (\ref{SG.cu.d}) describes a flow with unit speed, $|\gamma_t|=1$.  

The wave map equation (\ref{SG.cu.d}) and its conservation laws are invariant under the isometry group ${\rm U}(n+1,\mb{H})$ of $M=\mb{HP}^n$.

\subsection{Reductions and Curve Flows in $\mb{HP}^1$}

We remark that the mKdV and SG curve flows given by the geometric map equations (\ref{starstar}) and (\ref{SG.cu.d}) have a consistent reduction such that $\gamma(t,x)$ is a map into a submanifold $\mb{HP}^1\subset\mb{HP}^n$ or into $\mb{HP}^1$ itself when $n=1$. 
In the resulting curve flows, 
the components of the principal normal vector $N=\nabla_x\gamma_x$ along the curve $\gamma$ in a ${\rm U}(1,\mb{H})\times{\rm U}(n-1,\mb{H})$-parallel framing 
satisfy the scalar non-commutative mKdV and SG equations (\ref{red.mkdv}) and (\ref{red.sg}) respectively.

\section{Concluding Remarks}

Our derivation of quaternionic soliton equations (\ref{hier.ar}) 
and their Hamiltonian structure (\ref{hier.ar.1})--(\ref{hier.ar.2}) 
from geometric curve flows  (\ref{g.geom.d}) in $\mb{HP}^n$ 
can be reformulated entirely as an algebraic method at the level of 
the symmetric Lie algebra structure (\ref{rep.eigen.h})--(\ref{rep.eigen.m})
associated with the compact real symplectic group ${\rm Sp}(n)$ 
and the quaternionic unitary group ${\rm U}(n,\mb{H})$. 
Specifically, as shown in \cite{AncoJGP}, 
the Cartan structure equations (\ref{pull.1})--(\ref{pull.not.2}) 
for a framed curve flow in any symmetric space $M=G/H$
arise directly from the zero-curvature equation satisfied by 
the left-invariant $\mk{g}$-valued Maurer-Cartan 1-form $\omega_G$ 
on the Lie group $G$ viewed as a principal $H$-bundle over the manifold $M$. 
In this setting the pullback of $\omega_G$ by a local section $\psi:M \to G$
yields the $\mk{m}$-valued linear coframe $e$ 
and $\mk{h}$-valued linear connection 1-form $\omega$ on $M$ via
$e+\omega=\psi^*(\omega_G)$, where a change in $\psi\to \tilde\psi=\psi h$ 
corresponds to a local gauge transformation (\ref{gauge.tr}) 
on $e$ and $\omega$. 

Such a zero-curvature approach has been used in recent work \cite{AsadiSanders}
to derive a quaternionic mKdV equation for the variable 
$u=\omega_x=\omega\rfloor \gamma_x$,
based on the choice of a quaternionic connection matrix given by 
\begin{eqnarray*}
\conx_x=\begin{pmatrix}0&0&\tb{0}\\0&-\u&\tb{u}\\\tb{0}&-\ov{\tb{u}}^\t&\tb{0}\end{pmatrix}
\end{eqnarray*}
in terms of the imaginary scalar quaternion $\u$ and the $n-1$-component vector quaternion $\tb{u}$. 
This differs compared to our choice given by a parallel framing (\ref{mat.con})
and leads to a more complicated bi-Hamiltonian structure 
\cite{AsadiSanders}
$$ u_t = \tilde\Hop(\varpi^\perp) +\tilde\Nop(h^\perp),\quad
\varpi^\perp = \tilde\Jop(h^\perp)$$
where $\tilde\Hop$ and $\tilde\Jop$ are a compatible pair of
Hamiltonian cosymplectic and symplectic operators
and $\tilde\Nop$ is a non-trivial Nijenhuis operator \cite{Dorfman,Magri1980}, 
producing a different form than (\ref{mkdv.eq})--(\ref{mkdv.eq.v})
for the quaternionic scalar-vector mKdV equation 
arising from the flow defined by $h^\perp =u_x$. 
However, it was not shown in \cite{AsadiSanders} 
whether the above choice for the connection matrix 
can be achieved by a gauge transformation starting from an arbitrary form of
$u=\conx_x\in\mk{h}$. 

Building on work by one of us \cite{AsadiPhDthesis}, 
we will show in a forthcoming paper \cite{AncoAsadi}
that these bi-Hamiltonian operators are in fact related to our operators (\ref{ham.op}) and (\ref{sym.op}) 
by a Backlund transformation that can be interpreted as 
a Hasimoto gauge transformation corresponding to a change of framing. 
We will also derive Lax pairs for these quaternionic mKdV equations,
as well as the quaternionic SG equations (\ref{SG.eq})--(\ref{SG.eq.v}), 
by means of the zero-curvature equations (\ref{pull.1})--(\ref{pull.not.2}).

In addition, 
by expanding the imaginary scalar quaternion $\u$ 
and the ($n-1$ component) vector quaternion $\tb{u}$ 
in a quaternionic basis $\{1,i,j,k\}$, 
we plan to compare the resulting coupled integrable systems of
$4n-1$ ordinary (real-valued) scalar variables 
to other constructions of integrable multi-component mKdV and KdV type systems
known in the literature. 

Finally, in another direction, we also plan to explore the possibility of
deriving octonion soliton equations from bi-Hamiltonian geometric curve flows
in the octonion projective plane $\mb{OP}^2$ 
by adapting the zero-curvature/moving frame method used in the present paper.

\end{document}